\documentclass[aps,prl,showpacs,notitlepage,nofootinbib,superscriptaddress,floatfix,showkeys,twocolumn,preprintnumbers]{revtex4-1}

\usepackage{blindtext}
\usepackage{hyperref}
\usepackage{booktabs}
\usepackage{amsmath,amssymb}
\usepackage{float}
\usepackage{microtype}
\usepackage{graphicx}
\usepackage[caption=false]{subfig}
\usepackage{bm}
\usepackage{latexsym}
\usepackage{epsfig}
\usepackage{psfrag}
\usepackage{color}
\usepackage[dvipsnames]{xcolor}
\usepackage{subfig}
\usepackage[section]{placeins}
\usepackage{comment}

\def\e1i{\epsilon_{1\mathrm{i}}}

\allowdisplaybreaks[1]

\begin{document}

\title{Galactic cold clouds constraints on sub-GeV DM and asteroid-mass PBHs}

\author{Asier Salces Pérez}
\email{asier.salces@estudiante.uam.es}
\affiliation{Departamento de F\'{i}sica Te\'{o}rica, M-15, Universidad Aut\'{o}noma de Madrid, E-28049 Madrid, Spain}
\author{Pedro De la Torre Luque}\email{pedro.delatorre@uam.es}
\affiliation{Departamento de F\'{i}sica Te\'{o}rica, M-15, Universidad Aut\'{o}noma de Madrid, E-28049 Madrid, Spain}
\affiliation{Instituto de F\'{i}sica Te\'{o}rica UAM-CSIC, Universidad Aut\'{o}noma de Madrid, C/ Nicol\'{a}s Cabrera, 13-15, 28049 Madrid, Spain}

\smallskip
\begin{abstract}
We show that the ionization of molecular and atomic clouds provides a novel probe of dark matter scenarios producing low-energy $e^+e^-$ pairs through annihilation, decay, or Hawking evaporation of primordial black holes. We derive constraints on MeV-scale dark matter with masses between $\sim1$ and $100$ MeV, as well as on primordial black holes in the mass range $10^{14}$--$3\times10^{17}$g. By modeling the propagation of electrons and positrons inside these clouds, we show that uncertainties in charged-particle transport constitute the main limitation of this method. Nevertheless, for the most physically motivated propagation scenarios, the resulting constraints remain competitive with the strongest bounds currently available. We also identify the cloud properties that maximize the sensitivity to dark matter-induced ionization and discuss how larger samples of clouds, together with improved modeling of cosmic-ray ionization and cloud structure, could substantially enhance the reach of this technique. Our results establish molecular and atomic cloud ionization as a promising and complementary probe of sub-GeV dark matter and evaporating primordial black holes.
\end{abstract}
\maketitle

\section{Introduction}
\label{sec:Int}
The nature of dark matter (DM) remains one of the main open problems in astroparticle physics. Although its gravitational effects are well established over a wide range of astrophysical and cosmological scales \cite{Bertone:2004pz, Cirelli:2024DarkMatter, Clowe:2006eq, Planck:2018vyg, Rubin:1980zd}, no conclusive non-gravitational signal has been detected so far. This has motivated a coordinated effort involving collider experiments, direct detection searches, and indirect detection strategies to carry out a broad phenomenological exploration of a wide range of DM candidates and interaction mechanisms. In this context, indirect searches play a central role, as the presence of DM in astrophysical environments could lead to the production of Standard Model particles that are injected into the interstellar medium, giving rise to potentially observable signatures.

For many years, most experimental and theoretical efforts have focused on weakly interacting massive particles (WIMPs) \cite{Jungman:1995df}. However, the absence of a confirmed signal, together with the extensive regions of parameter space probed (particularly by direct detection experiments) has ruled out a large fraction of the viable WIMP parameter space \cite{LZ:2024zvo, XENON:2020fgj, Fermi-LAT:2015att, Conrad:2014tla}.
This has motivated the exploration of alternatives beyond the traditional paradigm. Among them, light, sub-GeV, DM has received increasing attention because it provides a well-motivated theoretical alternative in specific scenarios, such as resonant or asymmetric DM \cite{Balan:2024cmq, Battaglieri:2017aum, Cirelli:2024DarkMatter}, and encompasses several well-motivated DM candidates, including axions, axion-like particles, sterile neutrinos, and dark photons~\cite{Nguyen:2025tkl}. It has also been proposed as a possible explanation for several low-energy astrophysical anomalies \cite{laTorreLuquePedro:2024est, Huh:2007zw, DelaTorreLuque:2024fcc}. If DM lies in the MeV mass range and couples to electrons, its annihilation or decay can produce low-energy electron-positron pairs permeating the Galaxy. This occurs, for example, in models where light DM interacts with the Standard Model through a light vector or scalar mediator, allowing annihilation or decay channels into charged leptons~\cite{Boehm:2020wbt,Dutra:2018gmv}. Such scenarios lead to a phenomenology that differs from that of heavier DM candidates, since the injected particles mainly affect their environment by depositing their energy locally rather than by producing high energy emission. 

A different but equally well motivated possibility is that a fraction of the DM is made of primordial black holes (PBHs). In particular, PBHs in the lower end of asteroid-mass range are of special interest because they are light enough to emit Standard Model particles through Hawking radiation, while still being compatible with several existing constraints \cite{Carr:2026hot}. For PBH masses in the range $10^{14}$--$3\times10^{17}\,\mathrm{g}$, the Hawking spectrum includes electrons and positrons in the low-energy range. Therefore, both MeV particle DM and the lightest asteroid-mass PBHs could have a common phenomenological signature: the injection of low-energy charged particles into astrophysical environments. This opens the possibility of searching for their effects in systems where such particles can efficiently interact with the surrounding medium.

Cold atomic clouds and molecular clouds (MCs) provide particularly suitable environments for this purpose, as recently discussed in
Ref.~\cite{DelaTorreLuque:2025zjt}. Their cold gas allows low-energy electrons and positrons to efficiently deposit their energy through ionization of the dominant hydrogen component~\cite{Padovani:2009bj,Padovani:2013pha,Gabici:2022nac, Neufeld:2017}. Since the ionization rate can be inferred from suitable atomic and molecular tracers, it provides a direct observable to constrain additional exotic contributions
\cite{Indriolo:2012,Neufeld:2017}. 
 However, these regions are not uniform environments. Their ionization rates vary significantly from cloud to cloud, reflecting differences in CR fluxes, gas density, magnetic fields, turbulence, and transport conditions~\cite{Gabici:2022nac,Phan:2022iaq,Ng:2026rjn}.
The propagation of low-energy charged particles must therefore be modeled explicitly, since diffusion can allow them to escape before depositing a significant fraction of their energy.

In this work, we study the ionization signal induced by electrons and positrons produced by different DM scenarios in a set of cold-gas targets spanning both molecular and predominantly atomic environments. We consider annihilating and decaying MeV-scale particle DM, as well as evaporating PBHs in the lower end of the asteroid-mass range. For each case, we compute the injection spectrum of electrons and positrons and solve their transport inside the cloud using a diffusion-loss equation. The resulting stationary particle distribution is then used to calculate the ionization rate produced by the DM contribution. By requiring that the DM-induced ionization rate does not exceed the $2\sigma$ upper value of the ionization rate inferred in these clouds, we derive constraints on the annihilation cross section and decay lifetime of light particle DM. We also obtain bounds on the fraction of DM that can be made of asteroid-mass PBHs. Our results show that cold clouds can serve as a useful and complementary probe of DM models that inject low-energy electrons and positrons, offering the opportunity to explore different regions of the Galaxy using a single class of target.

\section{Ionization of molecular and atomic clouds}
\label{sec:Ionzrate}
MCs are among the coldest and densest regions of the interstellar medium. They are mainly composed of molecular hydrogen, with temperatures typically of order $T\sim 10-100\, {\rm K}$, and provide the sites where dense cores can collapse and form stars \cite{2006ARA&A..44..367S}. 
A similar class of clouds, typically characterized by lower gas densities but similarly low temperatures, consists of cold atomic-hydrogen clouds. Despite these low temperatures, molecular and atomic clouds are not completely neutral. Their ionization state is maintained by several possible agents, including ultraviolet radiation from nearby stars, X-rays, shocks, turbulence, supernova-driven CRs, and, in principle, any additional source of energetic particles able to penetrate the gas \cite{Meijerink:2010vv, Gabici:2022nac}. These different contributions are expected to leave different spatial and spectral signatures, depending on where the particles are produced and how efficiently they propagate inside the cloud.

For MCs, the relative importance of these mechanisms depends strongly on the column density of the cloud.\footnote{The column density is defined as the integral of the density along the line of sight,
$N_{\mathrm{H}_2}=\int_{\rm l.o.s.}n_{\mathrm{H}_2}\,{\rm d}l$.}
For diffuse MCs, with $N_{\mathrm{H}_2}\lesssim 10^{22}\,{\rm cm^{-2}}$, UV photons can still penetrate a significant fraction
of the gas. In contrast, at larger column densities, characteristic of dense or dark clouds \cite{2006ARA&A..44..367S}, the UV field is strongly attenuated towards the interior. In these regions, CRs are usually expected to be the dominant source of ionization \cite{1989ApJ...345..782M}, although the resulting ionization rate can still depend strongly on the gas density and
cloud structure. This CR ionization is fundamental for the chemistry of neutral gas, since it controls the abundance of molecular ions, contributes to the free-electron fraction, and can affect the thermal balance and evolution of the cloud.

A precise prediction of the CR ionization rate is not a trivial task. Early estimates for clouds mostly composed of molecular hydrogen found values spanning several orders of magnitude, ${\rm few}\times 10^{-18}\,{\rm s^{-1}}\lesssim\zeta^{\rm H_2} \lesssim10^{-15}\,{\rm s^{-1}}$ reflecting the poor knowledge in the low-energy CR spectrum \cite{1968ApJ...152..971S,2012ApJ...745...91I,Phan:2022iaq,Indriolo_2015}. This part of the spectrum is the most relevant one for ionization. For instance, the ionization cross section of atomic hydrogen peaks around $50\,{\rm eV}$ for CR electrons and around $10\,{\rm keV}$ for CR protons \cite{2009A&A...501..619P}. Therefore, a reliable calculation of the ionization rate, either $\zeta^{\mathrm{H}_2}$ in molecular gas or $\zeta^{\mathrm{H}}$ in predominantly atomic gas, requires knowledge of the spectra of CR electrons, protons and heavier nuclei well below the energies that are directly measured by current CR observations. This is observationally challenging. Measurements of low-energy CRs near Earth are affected by solar modulation and by the heliospheric magnetic field. Voyager~I and II have provided the most direct information outside the heliosphere, measuring the electron spectrum down to a few MeV \cite{Cummings:2016pdr,2019NatAs...3.1013S}, but even these energies are still far above the range where ionization is most efficient. Other probes, such as the Galactic synchrotron background, can provide complementary information on the low-energy electron population, but they depend on assumptions about the interstellar magnetic field and still require extrapolations to much lower energies \cite{Padovani:2018ypk,Orlando:2017mvd,Ginzburg:1965su}. As a result, the ionization rate expected from CRs inside dense cold neutral gas remains uncertain, and their propagation through these environments is even less well understood.

The ionization rate can be inferred observationally from suitable tracers observed through infrared or radio spectroscopic lines. These tracers are useful because their abundances depend on the free-electron population and therefore on the ionization rate. In diffuse molecular gas, $\mathrm{H}_3^+$ is particularly suitable due to its relatively well known chemistry \cite{2013ChSRv..42.7763I}. The basic formation chain begins with the ionization of molecular hydrogen,
\begin{equation}
\mathrm{H}_2 + {\rm CR}
\longrightarrow
\mathrm{H}_2^+ + e^- ,
\end{equation}
followed by
\begin{equation}
\mathrm{H}_2^+ + \mathrm{H}_2
\longrightarrow
\mathrm{H}_3^+ + \mathrm{H}.
\end{equation}
In denser clouds, $\mathrm{H}_3^+$ is efficiently destroyed through proton-hop reactions \cite{Gabici:2022nac}, further complicating the chemical network. Therefore, it is required to search for complementary tracers such as  $\mathrm{HCO}^+$, $\mathrm{DCO}^+$, CO, OH or HD \cite{1998ApJ...499..234C}. Analogous interactions can be observed in atomic clouds, where the ionization rate can instead be constrained using tracers like $\mathrm{ArH}^+$, $\mathrm{OH}^+$, and $\mathrm{H_2O}^+$~\cite{Neufeld:2017}.

\begin{table*}[t]
\centering
\caption{
Cloud targets used in this work. We list the adopted gas density $n_{\rm H}$, cloud size $R_{\rm cl}$, Galactocentric distance $R_{\rm gal}$, the observed or modelled ionization rate $\zeta_{\rm obs/model}$, and the reference upper value $\zeta_{\rm UL}$ used to constrain the exotic contribution. For the inner-Galaxy clouds, $\zeta_{\rm UL}$ is taken conservatively from twice the largest CR ionization rates of the corresponding CLOUDY models; these two targets are diffuse atomic clouds, so their rates are quoted per hydrogen atom, while those of the rest of targets are per H$_2$ molecule.}
\label{tab:clouds}
\begin{tabular}{lccccc}
\hline
Target
& $n_{\rm H}\,[{\rm cm^{-3}}]$
& $R_{\rm cl}\,[{\rm pc}]$
& $R_{\rm gal}\,[{\rm kpc}]$
& $\zeta_{\rm obs/model}\,[{\rm s^{-1}}]$
& $\zeta_{\rm UL}\,[{\rm s^{-1}}]$ \\
\hline
L1551-IR
& $3.2\times 10^{5}$
& $0.05$
& $8.277$
& $5.0\times 10^{-18}$
& $8.0\times 10^{-18}$ \\

HD 206165
& $1.0\times 10^{1}$
& $19.1$
& $8.277$
& $2.0\times 10^{-17}$
& $3.0\times 10^{-17}$ \\

DRAGON P8
& $4.6\times 10^{4}$
& $0.8$
& $4.5$
& $2.2\times 10^{-19}$
& $9.2\times 10^{-19}$ \\

G1.4-1.8+87
& $3.0\times 10^{-1}$
& $8.2$
& $0.44$
& $1\times 10^{-18}$
& $1\times 10^{-17}$ \\

G357.8-4.7-55
& $4.21\times 10^{-1}$
& $12.9$
& $1.1$
& $5\times 10^{-17}$
& $3.8\times 10^{-16}$ \\

\hline
\end{tabular}
\end{table*}
\subsection{Targeted Clouds}
\label{sec:target}

Observed ionization rates show a large cloud-to-cloud variation. If Galactic CRs were the only relevant ionizing source, one might expect $\zeta$ to decrease with increasing column density, since low-energy CRs should be progressively attenuated in denser gas. Some observations are compatible with this behaviour, but there is not a clear correlation yet \cite{Gabici:2019jvz,Gabici:2022nac}. In addition, several MCs exhibit ionization levels that are difficult to
reproduce with standard CR models \cite{Indriolo:2009tf,2018MNRAS.480.5167P}, leaving room for additional ionizing components. The Central Molecular Zone is an interesting example, since its large ionization rate is difficult to explain with conventional CRs alone
\cite{Ravikularaman:2024umo}, and MeV DM has been proposed as a possible exotic contribution \cite{DelaTorreLuque:2024fcc}.

Here, we describe a set of clouds selected to investigate how DM-induced ionization can be used to constrain DM self-annihilation and decay, as well as other particle injection scenarios, such as those associated with light PBH evaporation.
We consider a sample of clouds located at different positions throughout the Galaxy and spanning a wide range of physical properties to investigate which characteristics make them the most suitable targets for this type of search. Rather than providing an exhaustive catalog, the sample is designed to cover a broad range of gas densities, Galactocentric distances, and reference ionization rates. This enables us to assess how the sensitivity to DM-induced electrons and positrons depends on both the astrophysical properties of the clouds, their environment and the  DM density.

In particular, we consider:
\begin{itemize}
    \item \textbf{Local MCs:} We study the cases of L1551 (in particular, L1551-IR) and HD 206165, located in our local neighborhood. The advantage of using these clouds is that the DM density at their locations is much better known than in the Galactic Center region, and the ionization induced by CRs is better constrained thanks to local CR measurements. While L1551-IR is a dense core embedded in the L1551 MC, HD 206165 is an extended diffuse cloud. 
    
    \item \textbf{A dense subregion within the DRAGON molecular complex:} We also consider the particular case of the P8 subregion within the DRAGON cloud~\cite{Entekhabi:2022}. This target is especially interesting because it is located at an intermediate Galactocentric distance, where the uncertainty in the DM density is relatively small and comparable to that of the local DM density. In addition, the measured ionization rate is exceptionally low, owing to its high gas density and its location deep inside the larger MC, where the CR-induced ionization rate is likely suppressed.
    
    \item \textbf{Cold atomic clouds near the Galactic Center:} We study G1.4-1.8+87 and G357.8-4.7-55 as representative cold neutral targets located in regions where the DM density might be very large. Unlike the molecular-cloud targets discussed above, these systems are predominantly atomic, and the ionization rate is therefore evaluated under the assumption that hydrogen is primarily atomic. Interestingly, both clouds have similar density and size, but their inferred ionization rates differ significantly.
\end{itemize}

The specific properties of the selected targets are listed in Table~\ref{tab:clouds}. The parameters for L1551 and HD~206165 are taken from Refs.~\cite{1998ApJ...499..234C, 2010ApJ...718.1019Y, Indriolo:2025vzy}, while those of DRAGON~P8 from the analysis reported in Ref.~\cite{Entekhabi:2022}, for the Case 1 analysis (quoted as the most robust case). 
 In turn, the gas properties and \texttt{CLOUDY} ionization values of the inner-Galaxy clouds from Ref.~\cite{Bhoonah:2018gjb}. For each target, $\zeta_{\rm UL}$ is chosen as an upper limit for the exotic (DM) contribution. When an inferred ionization rate with its associated uncertainty is available, we take the corresponding $2\sigma$ upper value. For instance, for HD~206165 we use the analytical estimate $\zeta(\mathrm{H}_2)=(2.0\pm0.5)\times10^{-17}\,{\rm s^{-1}}$, which leads to $\zeta_{\rm UL}=3.0\times10^{-17}\,{\rm s^{-1}}$. In the case of the inner-Galaxy clouds, the quoted CR ionization rates correspond to the most extreme values inferred from the \texttt{CLOUDY} analysis~\cite{Ferland:2017Cloudy}, which was used to reproduce the observed properties of the clouds. In Ref.~\cite{Bhoonah:2018gjb}, the authors performed this analysis assuming three different cloud metallicities. Here, we adopt the metallicity yielding the highest ionization rate, as this choice leads to the most conservative constraint on the DM contribution. Furthermore, to set our constraint we define a conservative upper limit by imposing the ionization rate to be twice this maximum value.

We note, however, that G1.4-1.8+87 is subject to an additional observational uncertainty. 
Its originally inferred very low temperature relied on a narrow H$i$ spectral feature that was later questioned in Ref.~\cite{Farrar:2019qrv}; using the broader line instead implies a temperature of order $1.5\times10^{4}$~K. 
We therefore retain G1.4-1.8+87 only to illustrate the potential reach of such a cold inner-Galaxy target. We note, however, that the adopted upper-limit ionization rate is more than an order of magnitude higher than the average ionization rate measured in similar clouds in the Galaxy.

It is important to remark that the cloud size and density strongly affect the expected ionization profile. Low-density or small clouds are more easily penetrated by external UV/CRs, leading to a less attenuated and spatially flatter standard ionization profile. In contrast, dense and embedded regions can suppress the CR flux in their interiors, producing lower ionization rates and stronger gradients. This makes dense cores such as DRAGON~P8 particularly interesting, since their very low measured ionization rate leaves little room for an additional exotic component.

\section{Methodology}
\label{sec:methodology}

\subsection{Propagation of charged particles within cold clouds}

The transport of charged particles inside the cloud can be described by a diffusion-loss equation, in which spatial diffusion competes with continuous energy losses in the gas. In the energy range relevant for the DM candidates considered in this work, typically below $\sim 100\,{\rm MeV}$, the dominant loss channel in cold neutral gas is collisional ionization of the dominant hydrogen component \cite{evoli2017cosmic,Padovani:2009bj, schlickeiser2013cosmic}. This is due to the strong velocity dependence of ionization losses, which increase rapidly for non-relativistic particles, approximately as
\begin{equation}
\left|\frac{dE}{dt}\right|_{\rm ion}
\propto
\frac{n_{\rm H}}{\beta^2}.
\end{equation}
As a consequence, low-energy electrons and positrons will efficiently deposit their energy inside dense clouds.

The relative importance of losses and diffusion, however, depends on the properties of the cloud. In dense environments, ionization losses can dominate the evolution of sub-$100\,{\rm MeV}$ particles, making the cloud closer to a calorimetric target. In more diffuse clouds, the diffusion time may become shorter than the loss time, so that a significant fraction of the injected particles escapes before losing all its energy. At higher energies, other loss processes, such as bremsstrahlung, synchrotron emission, and inverse Compton scattering, become increasingly relevant. However, in the MeV energy range considered here, these processes remain subdominant compared to ionization losses in the cold gas, so that we can neglect them.
Furthermore, we also neglect advection and adiabatic losses, since typical turbulent bulk velocities in these objects are small, of order of a few ${\rm km\,s^{-1}}$, and their associated transport timescales are longer than those of diffusion or energy losses \cite{shi2024nuclear, Heiles:2002ju}.

The equilibrium timescale $t_{\rm eq} \simeq \min\left(t_{\rm loss},t_{\rm diff}\right)$
is typically much shorter than the lifetime of cold clouds, in the order of Myr \cite{clark2012long, mouschovias2006observational, Ward_2020}, for the parameter space considered here. We therefore work in the steady-state limit, $\frac{\partial N}{\partial t}=0$, 
where the transport equation becomes
\begin{equation}
-\nabla \cdot \left(D \nabla N\right) - \frac{\partial}{\partial p}\left(bN\right) =Q .
\label{eq:transport}
\end{equation}
Here $N(r,p)$ is the number density of electrons and positrons per unit momentum, $D(r,p)$ is the diffusion coefficient, $b(r,p)=-dp/dt$ is the momentum loss rate, and $Q(r,p)$ is the source term. The ionization losses are implemented as in \texttt{DRAGON2} (see Eq. (C.39) in \cite{evoli2017cosmic}).

The diffusion coefficient governing the propagation of these particles inside these clouds is not well constrained by observations. Although most Alfvén waves are expected to be damped in the dense environment of MCs, several studies have reported evidence for inhibited diffusion within these structures~\cite{Zweibel:1982, Gabici:2022nac}. Moreover, the propagation regime may vary throughout the cloud, with transitions between nearly ballistic and diffusive propagation depending on the cloud properties~\cite{Phan:2022iaq, 2018MNRAS.480.5167P}. Given these uncertainties, we adopt in our simulations for all targeted clouds a constant diffusion coefficient following the parametrization commonly used to describe the propagation of diffuse Galactic CRs~\cite{Luque_2024, Luque_MCMC, Weinrich_combined, Luque:2021joz, Luque_Ap, delaTorreLuque:2022vhm}:
\begin{equation}
    D = D_0 \beta^\eta \left(\frac{E}{E_0}\right)^\delta
\end{equation}
where $\beta$ is the speed of the particles in units of speed of light, and the other parameters are typically obtained from the observations of secondary-to-primary CRs. As our benchmark diffusion parameters, we adopt those obtained in Ref.~\cite{DelaTorreLuque:2024ozf}. However, we show the effect of changing this benchmark below.

Then, we consider three classes of DM source terms, depending on the process studied: annihilating particle DM, decaying particle DM, and evaporating PBHs. The corresponding source terms can be written as
\begin{equation}
Q(E,\mathbf{x}) =
\begin{cases}
\dfrac{\langle\sigma v\rangle}{2}
\left(\dfrac{\rho_\chi(\mathbf{x})}{m_\chi}\right)^2
\dfrac{dN^{\rm ann}}{dE},
& \text{annihilation}, \\[12pt]
\Gamma
\left(\dfrac{\rho_\chi(\mathbf{x})}{m_\chi}\right)
\dfrac{dN^{\rm dec}}{dE},
& \text{decay}, \\[12pt]
f_{\rm PBH}
\dfrac{\rho_{\rm DM}(\mathbf{x})}{M_{\rm PBH}}
\dfrac{d^2N_{\rm Hawk}}{dE\,dt},
& \text{PBHs}.
\end{cases}
\label{sourceterm}
\end{equation}
Here $\langle\sigma v\rangle$ is the velocity-averaged annihilation cross section, $\Gamma=1/\tau_\chi$ is the decay rate, $m_\chi$ is the particle DM mass, and $\rho_\chi(\mathbf{x})$ is the DM density in the cloud position. The quantities $dN^{\rm ann}/dE$ and $dN^{\rm dec}/dE$ denote the injected electron and positron spectra per annihilation or decay, which are delta functions peaking at $E\simeq m_\chi$ for annihilation
and $E\simeq m_\chi/2$ for decay. 
For PBH, $M_{\rm PBH}$ is the black hole mass, $f_{\rm PBH}$ is the fraction of DM in the form of PBHs, and $d^2N_{\rm Hawk}/dE\,dt$ is the Hawking emission rate per PBH, computed with the \texttt{BlackHawk} code \cite{arbey2019blackhawk}. We assume a monochromatic PBH mass function and consider non-rotating black holes. This provides a conservative benchmark, since rotation can modify and enhance Hawking emission through spin dependent greybody factors of modes aligned with the black hole angular momentum \cite{Auffinger:2022khh}. For each value of $M_{\rm PBH}$, the primary Hawking spectrum is computed with \texttt{BlackHawk}, including secondary emission through the \texttt{Hazma} option. The resulting electron and positron spectra are then used as the source term in the transport equation.

In general, both molecular and atomic clouds have non-uniform gas density distributions, which imply that both the diffusion coefficient and the energy loss rate vary throughout the cloud. In this work, however, we adopt a homogeneous cloud model to isolate the dominant physical effects and minimize the impact of additional astrophysical uncertainties. As discussed later, this simplification is not expected to significantly affect the average ionization rate predicted in the cloud.
Indeed, this approximation provides a controlled framework to study how DM-induced electrons and positrons propagate, lose energy, and contribute to the ionization rate.

\subsection{Ionization by dark matter}
Once the stationary distribution of electrons and positrons inside the cloud is obtained, we compute the corresponding hydrogen ionization rate. For the DM-induced contribution, this rate is given by
\begin{equation}
\zeta_{\rm DM}(r)
=
\int
J_{e^\pm}(r,E)\,
\sigma_{e,T}^{\rm ion}(E)\,
\left[1+\theta_e(E)\right]\,
{\rm d}E ,
\label{eq:zeta}
\end{equation}
where $J_{e^\pm}(r,E)$ is the differential particle flux of electrons and positrons inside the target, $\sigma^{\rm ion}_{e,T}$ is the electron-impact ionization cross section of the dominant hydrogen target $T$, and $\theta_e(E)$ accounts for secondary ionizations per primary ionization, following Ref.~\cite{Padovani:2009bj}. 
We adopt $T={\rm H}_2$ for the MC targets L1551-IR, HD 206165, and DRAGON P8, and $T={\rm H}$ for the cold neutral atomic-gas targets G1.4-1.8+87 and G357.8-4.7-55. 
For atomic hydrogen we use the relativistic RBEQ cross section of Kim, Santos \& Parente~\cite{PhysRevA.62.052710}, while for molecular hydrogen we use its $Q=1$ limit. The cross sections for the different target roughly differ by a factor of 2, while in both cases relativistic effects play an important role for probing high-energy (MeV) electron injections.
A detailed comparison between the different cross-section models and their impact on our results is presented in Appendix~\ref{sec:crosssectiondif}.

For atomic hydrogen we retain the secondary-ionization correction adopted for molecular hydrogen, since the mean energy per ion pair is nearly identical for the two targets \cite{Dalgarno_1999}. However, we note that the molecular case slightly underestimates the secondary contribution in atomic gas.

Given that, in this framework, DM injects particles approximately uniformly throughout the cloud (since the DM density varies negligibly over the spatial scales of this object), the resulting DM-induced ionization rate is expected to be nearly uniform throughout most of the cloud. Deviations from this behavior arise only near the cloud boundaries, where particle escape becomes significant. Higher-energy particles diffuse more efficiently and therefore have a shorter residence time inside the cloud, increasing the fraction of particles that escape before depositing all of their energy and consequently reducing the local ionization rate.

\begin{figure}[t]
    \centering
    \includegraphics[width=1\linewidth]{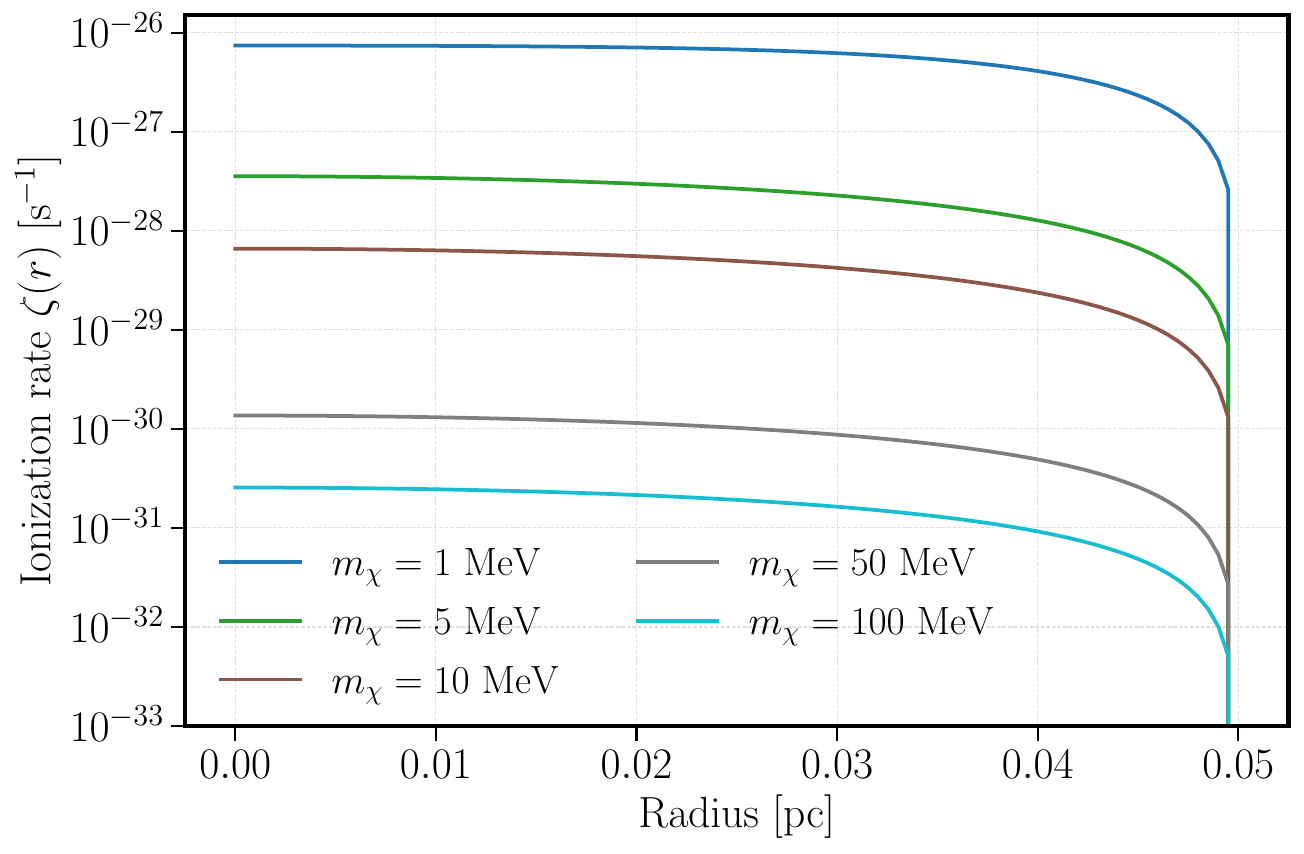}
    \includegraphics[width=1\linewidth]{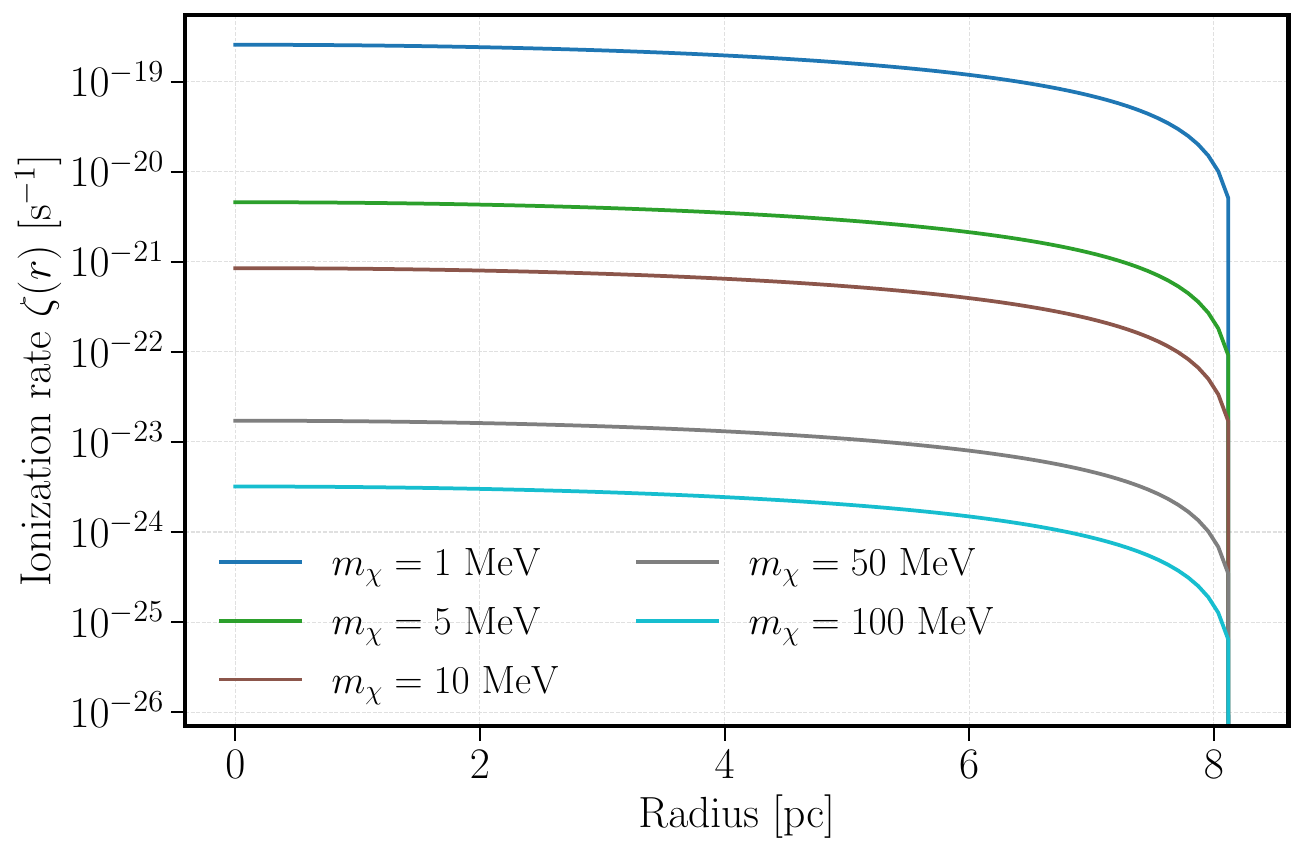}
    \caption{\textbf{Top panel:} Ionization rate profile inside the cloud in L1551 for five masses of annihilating DM with $\langle\sigma v\rangle = 1\times10^{-30} {\rm cm^{3}s^{-1}}$.
    \textbf{Bottom panel:} Ionization rate profile inside the cloud in G1.4-1.8+87 for five masses of annihilating DM with $\langle\sigma v\rangle = 1\times10^{-30} {\rm cm^{3}s^{-1}}$}
    \label{fig:NinOptimistic}
\end{figure}

This behavior is illustrated in the ionization profiles shown in Fig.~\ref{fig:NinOptimistic}, where we present the case of annihilating DM with a cross-section $\langle\sigma v\rangle =10^{-30} {\rm cm^{3}s^{-1}}$ for the local MC L1551-IR and the inner-cloud target G1.4$-$1.8+87, assuming the canonical diffusion coefficient $D_0=3\times10^{28}\,{\rm cm^2\,s^{-1}}$. These two clouds provide representative examples of environments with different sizes, gas densities, and DM densities, allowing us to illustrate how the spatial distribution of the ionization rate depends on both particle propagation and the cloud properties. Corresponding results for PBHs and decaying DM are presented in Appendix~\ref{sec:Setups}.

We also investigate the dependence of these profiles on the normalization of the diffusion coefficient. As expected, increasing the diffusion coefficient enhances particle escape by reducing the residence time within the cloud. Consequently, the average ionization rate decreases, and the suppression of the ionization rate near the cloud boundaries becomes more pronounced.

\begin{figure}[htb!]
    \centering
    \includegraphics[width=1\linewidth]{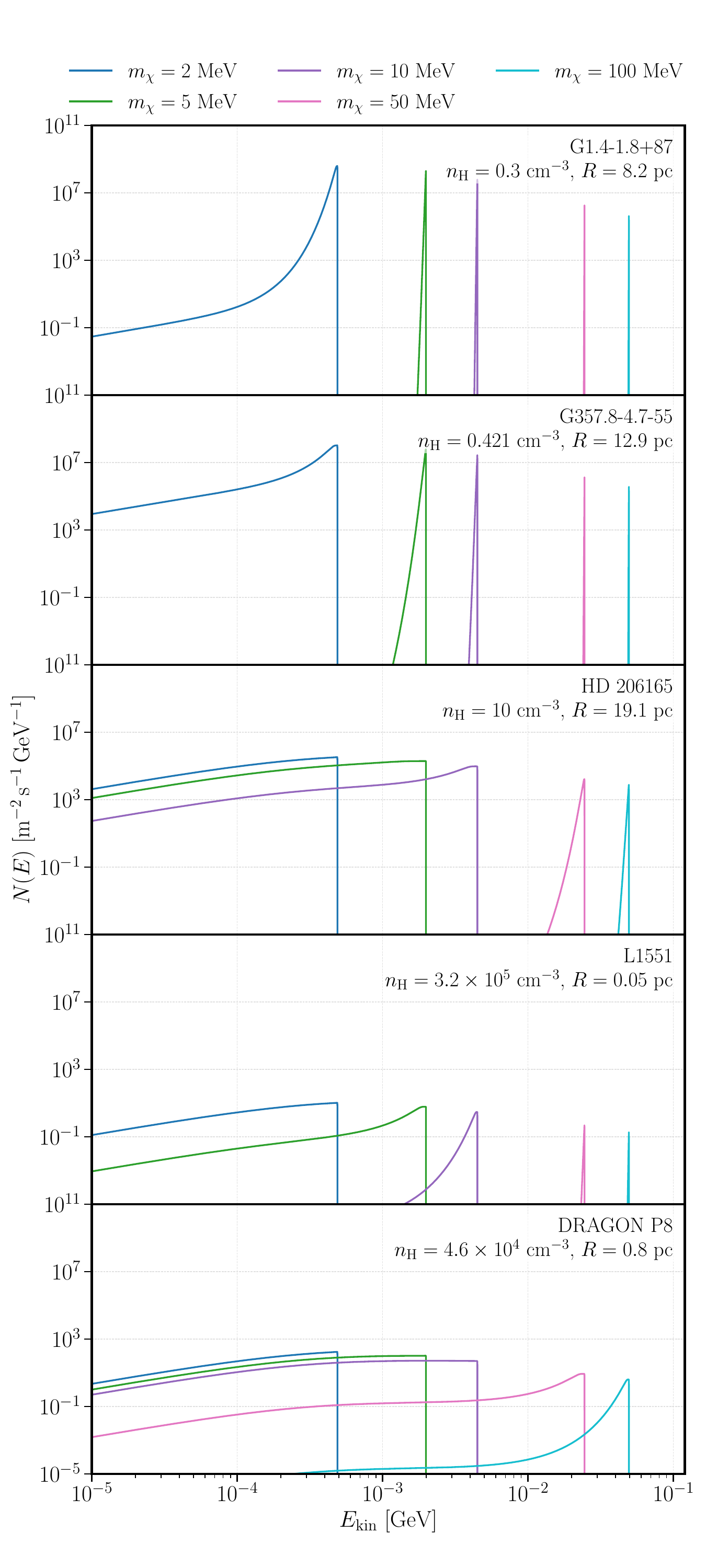}
  \caption{
    Propagated electron and positron spectra inside the targets for
    decaying DM into $e^+e^-$. Each panel corresponds to a different cloud, with
    the adopted gas density $n_{\rm H}$ and radius $R$ indicated in the plot. The
    different colors show the stationary spectra obtained for various DM masses for a fixed DM lifetime of $\tau = 1\times10^{26}{\rm s}$ and
    diffusion setup with $D_0=3\times10^{28}\,{\rm cm^2\,s^{-1}}$. The peaks correspond to the monoenergetic injection energy, while the low-energy tail is produced by energy losses during propagation.}
    \label{fig:multdecpectra}
\end{figure}

An interesting aspect to consider is the effect of propagation on the resulting particle spectrum, shown in Fig.~\ref{fig:multdecpectra} for all target clouds and for the case of decaying DM with a lifetime $\tau = 1\cdot10^{26}{\rm s}$. For some DM masses (towards heavy mass), the resulting electron-positron spectrum closely resembles the injection spectrum (a Dirac delta function), as the particles are injected at energies for which diffusion dominates their propagation. At lower energies, however, the spectra develop extended tails toward lower energies due to the increasing impact of energy losses. This interplay between propagation and energy losses depends not only on the particle energy but also on the physical properties of the cloud, such as its size and density. In particular, denser clouds enhance the effect of energy losses (as illustrated by the DRAGON P8 case), while larger clouds increase the residence time of particles within the molecular environment, leading to more significant energy degradation before escape.
A similar figure can be found in the Appendix~\ref{sec:Setups} for the case of evaporating PBHs and annihilating DM, in Fig.~\ref{fig:multicloud_spectra}.

\begin{figure}[h]
    \centering
    \includegraphics[width=1\linewidth]{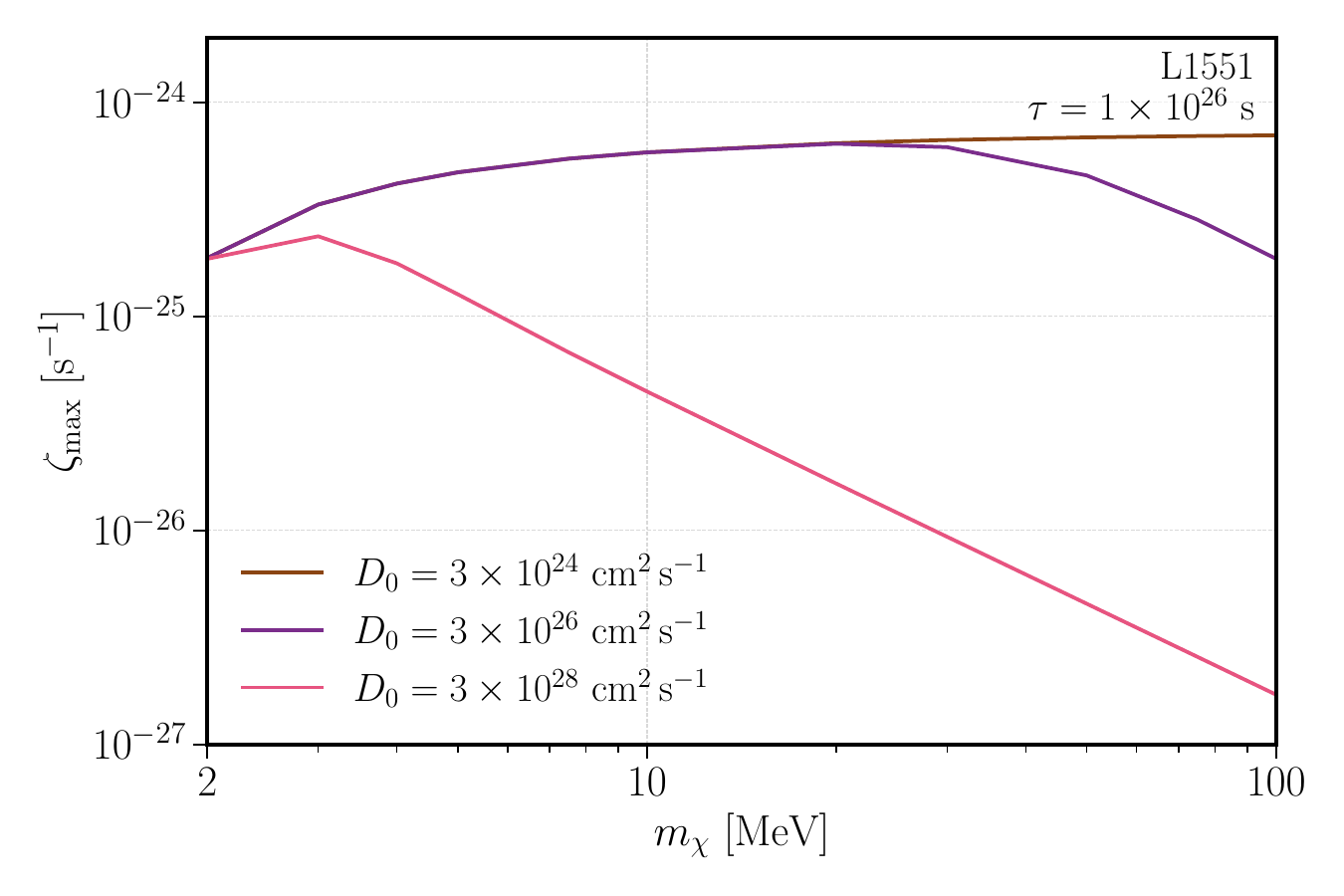}
    \includegraphics[width=1\linewidth]{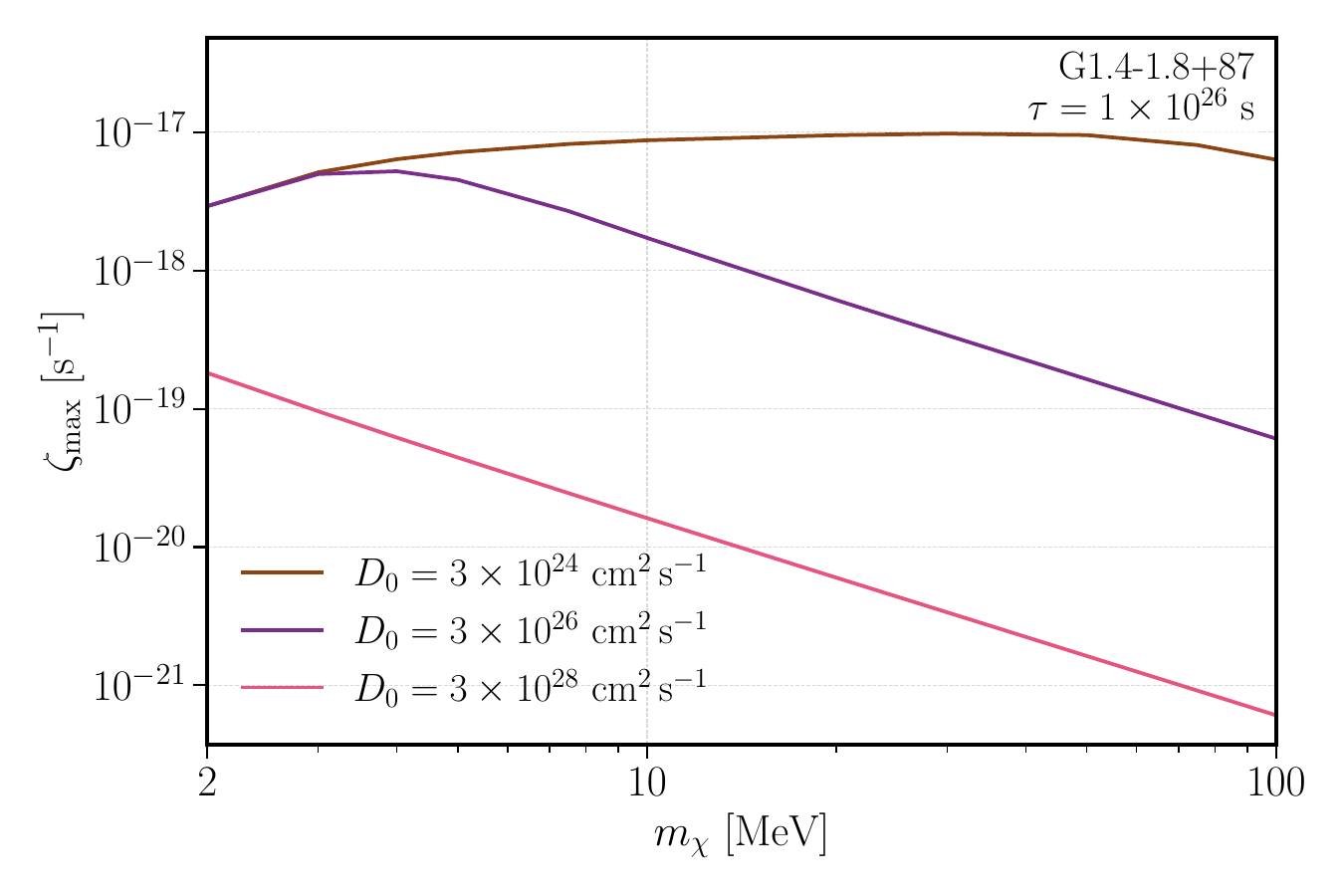}
    \caption{Ionization rate induced by decaying DM as a function of its mass for L1551-IR \textbf{(top)} and G1.4-1.8+87 \textbf{(bottom)}, assuming $\tau=1\times10^{26}\,{\rm s}$. The different curves correspond to three diffusion normalizations, $D_0=3\times10^{24}$, $3\times10^{26}$ and $3\times10^{28}\,{\rm cm^2\,s^{-1}}$. Suppressed diffusion leads to larger ionization rates, especially in the more diffuse G1.4-1.8+87 cloud.}
    \label{fig:ionmassdecay}
    \end{figure}

Finally, we show in Fig.~\ref{fig:ionmassdecay} the average ionization rate expected from DM decay in L1551 and G1.4-1.8+87 as a function of the DM mass. The different curves correspond to different normalizations of the diffusion coefficient and illustrate the transition between loss-dominated and escape-dominated propagation as a function of the DM mass.
At low DM masses, where the injected particle flux is high but the electrons and positrons have relatively low energies, the regime is loss-dominated. In this regime, most of the injected particles lose their energy within the cloud before escaping, making the ionization rate largely insensitive to the diffusion coefficient. This phenomenon is clear, for example, in G1.4$-$1.8+87, where the curves corresponding to the two lowest diffusion normalizations begin at nearly the same ionization rate. 

As the DM mass increases, the injected particles become more energetic and their energy-loss time increases relative to their escape time. The system therefore transitions smoothly from the loss-dominated to the escape-dominated regime. This transition is reflected in a change in the slope of the ionization rate: when energy losses dominate, the ionization rate decreases only slowly with the DM mass, whereas in the escape-dominated regime an increasing fraction of particles leaves the cloud before depositing their energy. Consequently, the ionization rate evolves as the particle injection rate, scaling as $1/m_{\chi}$ for decaying DM (and PBH evaporation) and $1/m_{\chi}^2$ for annihilating DM.

This transition crucially depends on the cloud properties. In L1551, the transition from the loss dominated regime to the escape dominated regime is more gradual and occurs at larger masses due to its larger density ($n_{\rm H}=3.2\cdot10^5\,{\rm cm^{-3}}$, compared to $n_{\rm H}=0.3\,{\rm cm^{-3}}$ in G1.4-1.8+87). The relevant behaviour is therefore controlled by the interplay between the cloud size, which determines the escape time, and the gas density, which controls the energy-loss time. For very suppressed diffusion, most of the mass range remains loss dominated, and the ionization rate therefore shows a much flatter dependence on $m_\chi$.

\subsection{Ionization by diffuse cosmic rays}
\label{sec:LISIon}

So far, we have not discussed the ionization produced by diffuse CR protons, which is expected to be a dominant ionization source in the interstellar medium and represents an unavoidable background contribution in cold clouds. We model this component using the same diffusion-loss framework described above in Eq.~\ref{eq:transport}, but with a different physical setup. In this case, we impose the external CR spectrum as the cloud boundary and propagate it inward.

To do this, we adopt a conservative approach and model the low-energy CR spectrum in the Galaxy as being uniform and given by the local interstellar spectrum (LIS) that reproduces local CR observations. In particular, we adopt the best-fit model of \cite{2015ApJ...815..119V}
\begin{equation}
    j_\mathrm{LIS}(E_k) = 2.70\,\frac{E_k^{1.12}}{\beta^2}\left(\frac{E_k + 0.67}{1.67}\right)^{-3.93},
\end{equation}
in units of $\mathrm{m^{-2}\,s^{-1}\,sr^{-1}\,MeV^{-1}}$, where $E_k$ is the proton kinetic energy in GeV, fitted to Voyager~1 \& 2, PAMELA, and AMS-02 data up to 100~GeV. 

The propagated proton spectrum is then converted into an ionization rate using the same expression as in Eq.~\ref{eq:zeta}, but replacing the electron and positron flux by the propagated proton flux, and using the effective proton ionization cross section of $\mathrm{H}_2$. For these cross sections, we adopt the parametrization of Ref.~\cite{2009A&A...501..619P}, which includes primary ionization, electron capture, and secondary ionizations. For the two atomic clouds we convert the resulting rate to an atomic-hydrogen ionization through the useful approximation $\zeta^{\mathrm{H}} = (1.5/2.3)\zeta^{\mathrm{H_2}}$ \cite{2009A&A...501..619P, Indriolo:2009tf, 1974ApJ...193...73G}. The contribution of heavier CR nuclei is included through
\begin{equation}
    \zeta_{p+Z}(r) = (1+\eta)\,\zeta_p(r),
\end{equation}
with $\eta=0.51$ \cite{2009A&A...501..619P, Indriolo:2009tf, Anders:1989zg}. Given that the LIS spectrum is expected to provide a good description of the CR flux in the local environment, we mainly focus on the standard CR-induced ionization rate in the local clouds L1551 and HD~206165. As mentioned earlier, in the case of L1551, the inferred ionization rate was measured in the dense core L1551-IR rather than to the full extended cloud.

\begin{figure}
    \centering
    \includegraphics[width=1\linewidth]{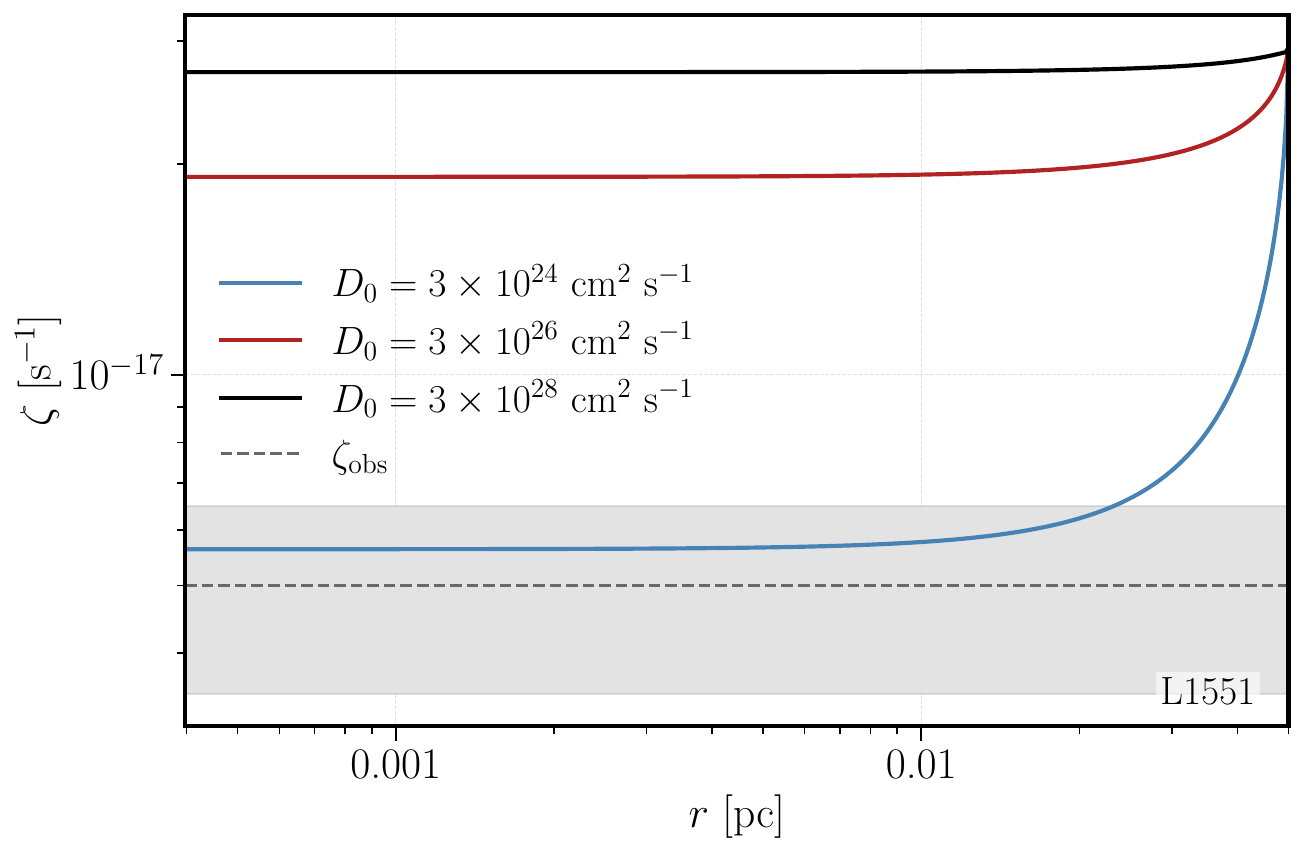}
    \includegraphics[width=1\linewidth]{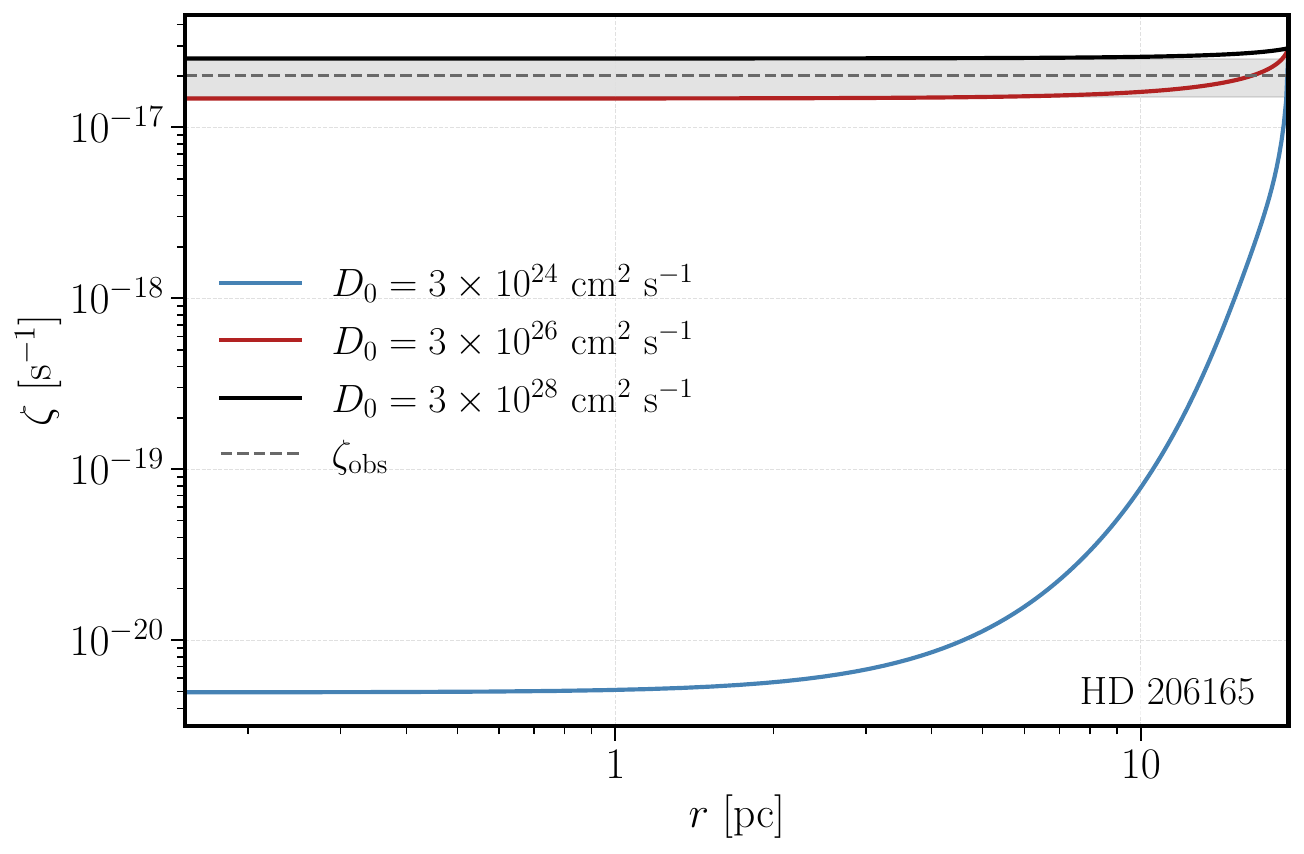}
    \caption{LIS-induced ionization profiles for the two local MC targets.
    The top panel shows the L1551-IR core, while the bottom panel shows
    HD~206165. The colored solid lines show the ionization profiles obtained for
    three diffusion normalizations, $D_0=3\times10^{24}$, $3\times10^{26}$ and
    $3\times10^{28}\,{\rm cm^2\,s^{-1}}$. The grey dashed line indicates the
    inferred ionization rate, $\zeta_{\rm obs}$, and the grey shaded band shows the
    corresponding observational uncertainty. The comparison illustrates that the
    very low ionization rate inferred in L1551-IR requires a strongly suppressed
    diffusion coefficient, whereas HD~206165 is compatible with a less suppressed
    transport regime.}
    \label{fig:LISioniz}
\end{figure}

The resulting ionization profiles are shown in Fig.~\ref{fig:LISioniz}. The predicted CR-induced ionization rate depends strongly on the assumed diffusion coefficient, with lower diffusion coefficients leading to lower ionization rates throughout the clouds.
For L1551, both the standard Galactic diffusion coefficient, $D_0=3\times10^{28}\,{\rm cm^2\,s^{-1}}$, and the moderately suppressed case, $D_0=3\times10^{26}\,{\rm cm^2\,s^{-1}}$, predict ionization rates that remain well above the adopted reference value, $\zeta_{\rm ref}=5\times10^{-18}\,{\rm s^{-1}}$, across the entire cloud. This suggests that CR propagation in L1551 is significantly more suppressed than expected from the average Galactic value.

The situation is less restrictive for HD~206165. The standard Galactic diffusion coefficient predicts an ionization rate only slightly above the inferred value, while the moderately suppressed case provides a closer match. However, this apparent agreement with the standard Galactic diffusion coefficient is likely optimistic. Owing to the diffuse nature of the cloud, additional ionization sources may contribute to the measured ionization rate. Accounting for these contributions would reduce the fraction of the ionization attributable to CRs, thereby favoring a lower effective diffusion coefficient.

Only for a much smaller normalization, $D_0=3\times10^{24}\,{\rm cm^2\,s^{-1}}$, does the predicted ionization become compatible with the inferred values in the L1551-IR cloud core. In this regime the ionization is strongly suppressed in the inner regions and rises towards the cloud boundary, showing that low-energy CRs mainly ionize the outer layers and do not efficiently penetrate the whole cloud. This conclusion should be treated with some caution, since the ionization measurement refers to the compact core, which is embedded in the larger L1551 cloud.  
We find that a very similar diffusion coefficient is still required when taking into account the full cloud structure, although the level of suppression is slightly weaker than considering an L1551-IR as an isolated core (see Fig.~\ref{fig:corepluscloud} in Appendix~\ref{sec:Setups}). This modest difference is expected, as both the core-only (0.05 pc) and core-plus-cloud (1 pc) configurations probe relatively small spatial scales. However, the latter result depends on the poorly constrained transition between the dense core and the surrounding diffusion envelope and the cloud properties. For this reason, we adopt the core-only calculation as our benchmark model for suppressed diffusion.

This level of suppression is much stronger than that inferred in recent multi-wavelength studies of CR transport in nearby MC substructures. For instance, Ref.~\cite{Ng:2026rjn} studied CR transport in the Taurus MC and embedded clumps using $\gamma$-ray, X-ray and ionization diagnostics, considering ballistic, diffusive and two-zone transport models. Their fits favour diffusion coefficients suppressed by at least $\gtrsim 2$ orders of magnitude relative to the canonical ISM value, with the most attenuated clumps requiring suppression approaching a factor $\sim 300$ at $1\,{\rm GeV}$. However, in our local cloud calculation even the intermediate value $D_0=3\times10^{26}\,{\rm cm^2\,s^{-1}}$, comparable to such a suppression, would still overproduce the observed ionization in L1551. Avoiding this requires $D_0$ to be reduced to values of order a few $10^{24}\,{\rm cm^2\,s^{-1}}$, or equivalently an additional mechanism that prevents low-energy CRs from penetrating efficiently into the gas. Another example where such a suppression of CR penetration has been suggested is the Central Molecular Zone, where a "barrier" for the Galactic CR sea has been inferred in Ref.~\cite{Huang:2021cmz}

Therefore, also adopting the suppressed diffusion normalization of $D_0=3\times10^{24}\,{\rm cm^2\,s^{-1}}$ as a second benchmark allows us to bracket the uncertainties in the predicted ionization profiles in the rest of the analysis. 
In this way, the standard Galactic value and the L1551-IR suppressed diffusion value provide two physically motivated limiting cases: efficient CR penetration and strong CR suppression. The true ionization signal is indeed expected to lie within this range for the typical conditions of the clouds. We use it to define the uncertainty bands in the DM and PBH constraints to show the potential and limitations of this analysis.

\begin{figure}[th!]
    \centering
    \includegraphics[width=1\linewidth]{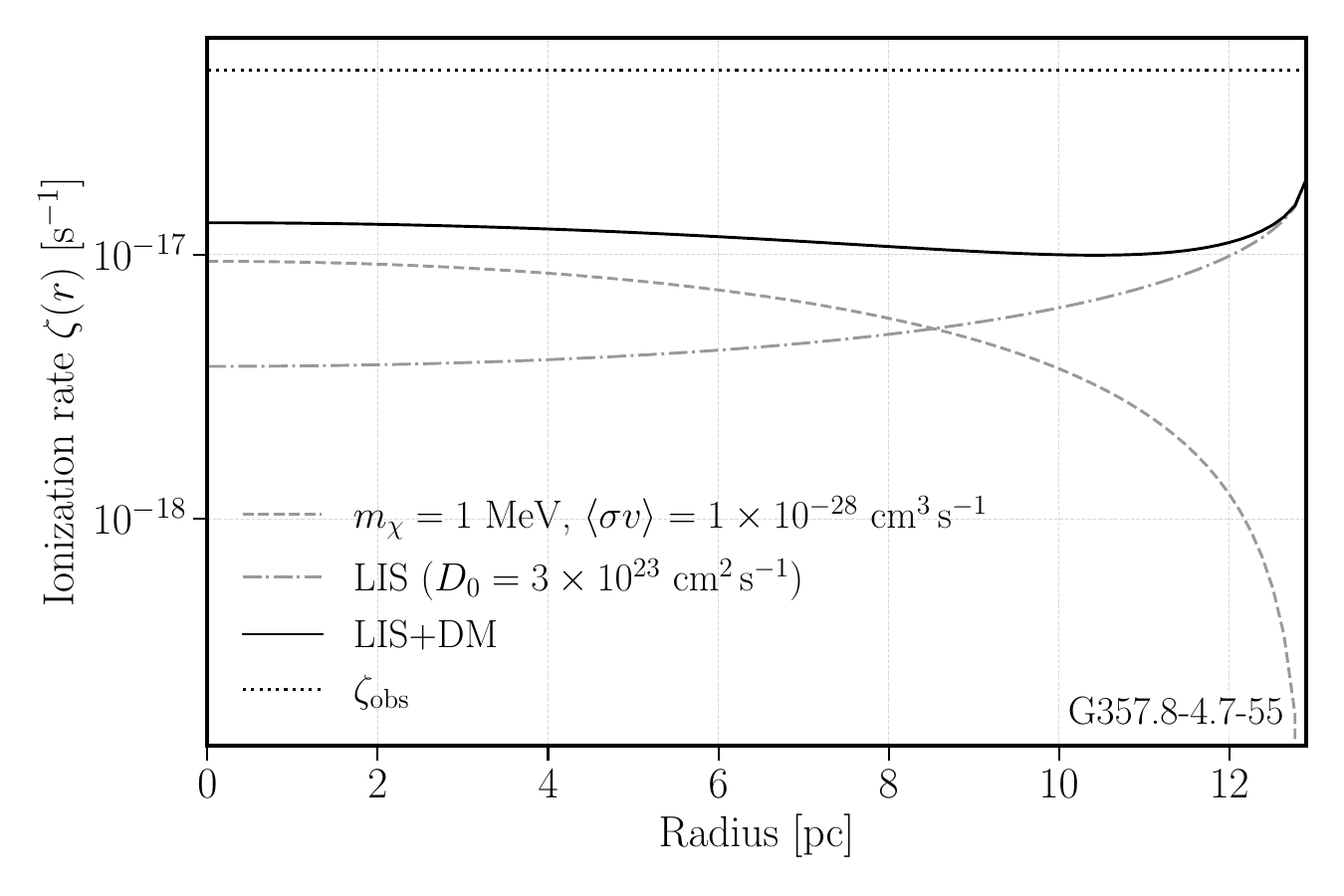}
    \caption{Comparison between the ionization profiles induced by the LIS CR component and by annihilating DM in G357.8-4.7-55. The dashed line shows the DM-induced ionization profile, while the dot-dashed line shows the LIS-induced contribution for a strongly suppressed-diffusion benchmark. The solid line corresponds to the total ionization rate, obtained as the sum of the LIS and DM contributions. The dotted horizontal line indicates the adopted reference ionization rate. The parameters are chosen so that the two contributions are of comparable magnitude, in order to highlight their different radial behaviour.}
    \label{dmvslis}
\end{figure}

We also compare the radial profiles expected from the LIS and DM components in Fig.~\ref{dmvslis}. Both the DM and LIS parameters are chosen so that the two contributions are of comparable magnitude, allowing us to illustrate how each component would shape the total ionization profile. The LIS-induced ionization is larger close to the cloud boundary, as expected for CRs entering the cloud from the outside and losing energy as they propagate inward. By contrast, the DM contribution is generated throughout the cloud volume and therefore produces a smoother profile, less concentrated near the outer layers as particles scape. This comparison shows that an exotic contribution would not simply act as an overall rescaling of the standard CR ionization rate. Instead, it could change the radial dependence of the ionization profile, especially in dense regions where the LIS contribution is suppressed by inefficient CR penetration.

\section{Results}

In this section we derive constraints on sub-GeV DM and asteroid-mass PBHs from the ionization rates of the cold cloud targets described above. For each model, we require the DM-induced contribution to match the reference ionization rate adopted for each cloud. It is important to remark that the LIS-induced ionization is not added when deriving the limits. This choice is conservative, since including the standard CR background would reduce the ionization budget available for a DM component. Therefore, the LIS calculation is used only to motivate a suppressed-diffusion benchmark, which allows us to show the impact of transport uncertainties on the DM constraints and its potential when this value is reduced (as expected).

As our reference case, we adopt a standard NFW profile for the Galactic DM density distribution~\cite{Navarro:1995iw}. The impact of uncertainties in the DM density profile is discussed in Appendix~\ref{sec:profileunc}. For the local clouds, we adopt a reference local DM density of $\rho_\odot = 0.4\,{\rm GeV\,cm^{-3}}$~\cite{Benito_2019} and vary it within the range $\rho_\odot=0.2$--$1\,{\rm GeV\,cm^{-3}}$ to assess the impact of this uncertainty. For clouds located towards the inner Galaxy, we consider the Moore~\cite{Moore:1999nt} and Burkert~\cite{Burkert:1995yz} profiles as optimistic and pessimistic cases, respectively.
   
\subsection{Constraints on sub-GeV DM}

\begin{figure}[t]
    \centering
     \includegraphics[width=1\linewidth]{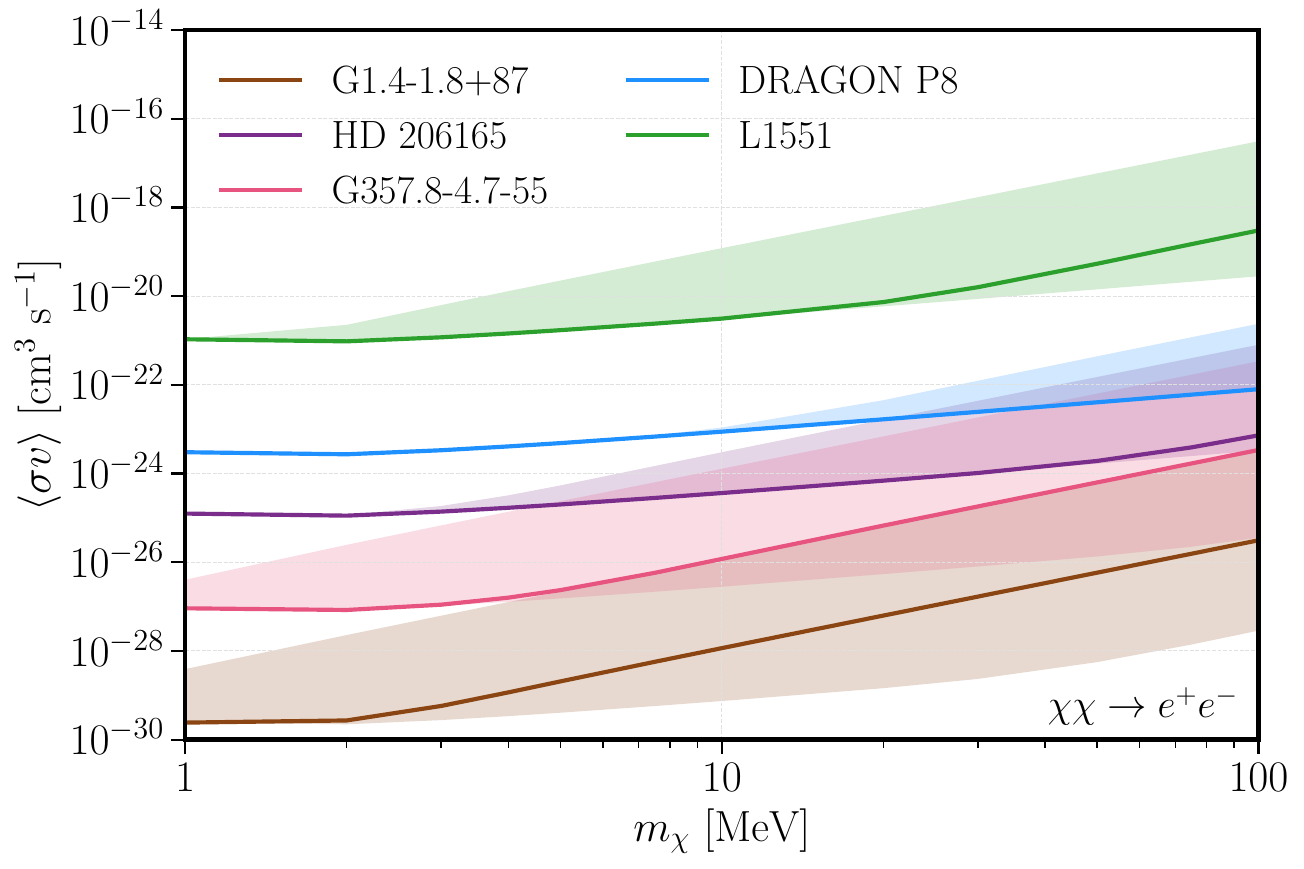}
     \centering
    \includegraphics[width=1\linewidth]{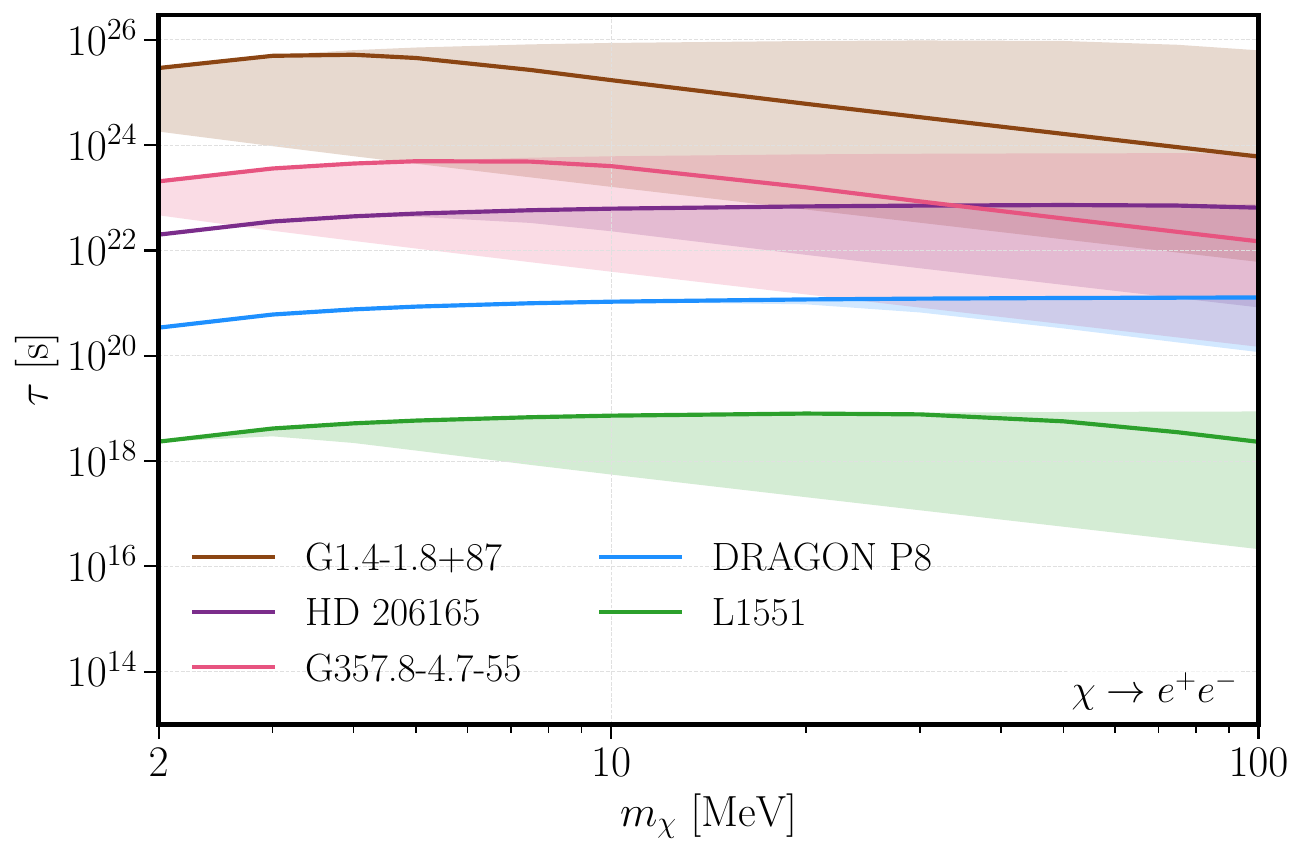}
    \caption{Constraints on annihilating DM \textbf{(top)} and decaying DM \textbf{(bottom)} for the cold clouds considered in this work. The shaded bands bracket the effect of varying the diffusion normalization between the standard Galactic value and the suppressed-diffusion benchmark, while the solid lines show the intermediate case $D_0=3\times10^{26}\,{\rm cm^2\,s^{-1}}$.}
    \label{constanndecaybands}
    \end{figure}

We first consider light DM particles in the range of $1-100 \,{\rm MeV}$ annihilating and decaying into $e^+e^-$ pairs, being agnostic of their cosmological origin or type of particle. The resulting constraints are shown in Fig.~\ref{constanndecaybands} for the clouds considered in this work. For each target, the colored band shows the range of constraints obtained between the standard Galactic diffusion normalization $D_0=3\times10^{28}\,{\rm cm^2\,s^{-1}}$, and the strongly suppressed value we find from the L1551-IR core, $D_0=3\times10^{24}\,{\rm cm^2\,s^{-1}}$. The solid line shows the intermediate case, $D_0=3\times10^{26}\,{\rm cm^2\,s^{-1}}$, which we adopt as an intermediate and reliable case, accounting for both inhibited transport while avoiding the extreme suppression by the LIS analysis alone.

The effect of diffusion is notable in these plots. A smaller diffusion coefficient increases the residence time of the injected leptons inside the cloud, allowing them to lose a larger fraction of energy through ionization (i.e. leading to more energy exchange between DM and the cloud). Consequently, suppressed diffusion leads to stronger constraints. On the contrary, for larger diffusion coefficients, particle escape becomes more efficient. 
We notice that the size of the band also depends on the cloud properties. Diffusion has a larger impact in clouds where particle escape is easier, that is in smaller and more diffuse targets. In larger or denser clouds, particles remain confined for longer times and can lose a bigger fraction of their energy before escaping, so the limits are less sensitive to this parameter. An extreme example is DRAGON P8, which is both very dense and sufficiently large. In this case, energy losses dominate over escape over most of the parameter space, and the constraints obtained with the canonical and suppressed diffusion benchmarks are nearly identical. As a result, a really narrow uncertainty band is obtained for this target. This is also visible when comparing HD 206165 and G1.4-1.8+87 behaviors. Over all the mass range, HD 206165 remains close to the loss-dominated regime because of its large size, and the corresponding band matches the solid line. By contrast, G1.4-1.8+87 is both smaller and less dense, so escape is already important at low masses and the resulting band is much wider.

For annihilating DM, the strongest limits are always obtained from the inner-Galaxy clouds. This is expected since the annihilation source term in Eq.~\ref{sourceterm} is proportional to $\rho_\chi^2$ making the result very sensitive to the DM density at the cloud position. The bounds generally become weaker at larger DM masses because the number of injected particles decreases and higher-energy particles escape more efficiently. However, this trend can be partially compensated by the
relativistic flattening and subsequent rise of the ionization cross section, so that some constraints remain nearly flat or even mildly strengthen over a
limited mass range (see Appendix~\ref{sec:crosssectiondif}).

For decaying DM, the signal scales only linearly with the DM density. The difference between targets is therefore less pronounced than in the annihilation case, and the cloud properties become comparatively more relevant. Interestingly, HD 206165 can provide stronger limits at high DM masses than G357.8-4.7-55, despite having a smaller DM density. This is due to its larger size and higher gas density, which increase the residence time and the ionization efficiency of the injected particles. However, when suppressed diffusion is considered, the enhancement of confinement is larger for G357.8-4.7-55, so its upper edge can again overtake the HD~206165 bound. This illustrates that the strength of the limits is controlled not only by the DM density, but also by the interplay between cloud size, gas density, and diffusion.

Finally, we compare our best limits with existing bounds in Fig~\ref{fig:constcompare}. For annihilating DM, our limits remain weaker than the strongest existing constraints over most of the mass range, in particular those from the CMB~\cite{Slatyer2016CMB}, Leo~T~\cite{Wadekar_2022} and the Galactic 511 keV line morphology~\cite{DelaTorreLuque2024}. Nevertheless, at low masses, the G1.4-1.8+87 cloud can reach a similar level to both CMB and Leo~T constraints, significantly improving upon previous X-ray limits~\cite{Balaji:2025afr}. For decaying DM, our bounds become more competitive. The strongest limit is again obtained for G1.4-1.8+87, which improves over the CMB and gamma-ray limits~\cite{Essig:2013goa} at low masses, although it gets weaker at larger masses, when particle escape becomes more important. The G357.8-4.7-55 constraint is weaker, but still lies in a similar range to the CMB and gamma-ray results. However, they remain below the most stringent bounds from Voyager~1~\cite{DelaTorreLuque2024} and the Galactic 511 keV line. For illustration purposes, we do not include the Leo~T gas-heating
constraints~\cite{Wadekar_2022}. We nevertheless highlight the recent
H$\alpha$ limits from Leo~T~\cite{Leane:2025ohu}, which provide interesting constraints in the same mass range, reaching $\langle\sigma v\rangle\sim
10^{-30}\,{\rm cm^3\,s^{-1}}$ for annihilation and lifetimes of order
$10^{26}\,{\rm s}$ for decaying DM with future projections extending further.

\subsection{Constraints on PBHs}

\begin{figure}[t]
    \centering
    \includegraphics[width=1\linewidth]{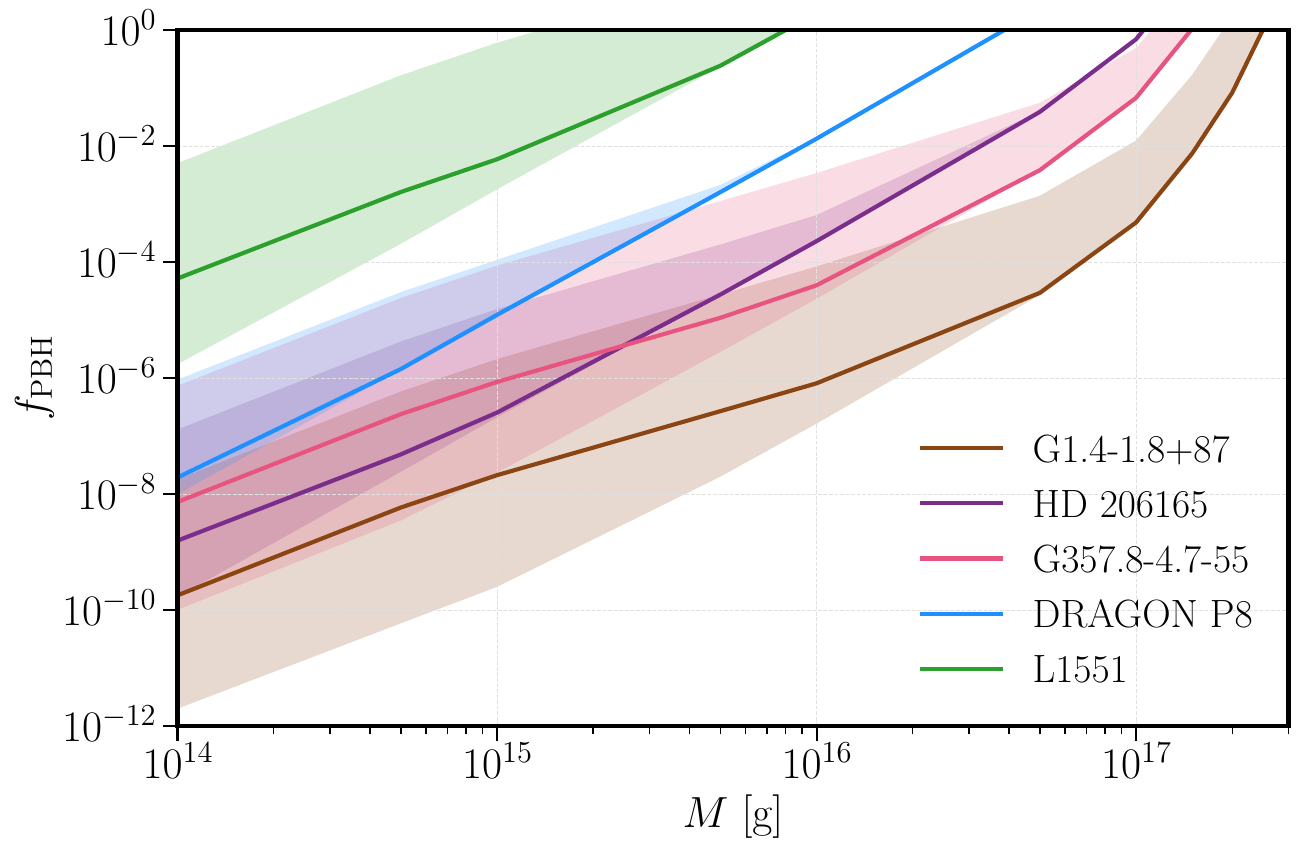}
    \caption{Constraints on the fraction of PBHs in the form of DM for the clouds considered in this work. The shaded bands bracket the effect of varying the diffusion normalization between the standard Galactic value and the suppressed-diffusion benchmark, while the solid lines show the intermediate case $D_0=3\times10^{26}\,{\rm cm^2\,s^{-1}}$.}
    \label{constpbhband}
    \end{figure}

We also derive bounds on the fraction of DM in the form of evaporating PBHs, assuming a monochromatic mass function and non-rotating PBHs, as a conservative common assumption. The resulting limits are shown in Fig.~\ref{constpbhband}. As in the particle-DM case, the bands represent the variation of the diffusion normalization. 

Unlike the particle-DM cases, PBH evaporation does not inject a monoenergetic electron-positron population, but a continuous Hawking spectrum. The constraints are stronger at lower PBH masses, because of a higher injection spectrum (larger $\rho/m_{PBH}$). Moreover, in this regime, the Hawking temperature is larger since it scales inversely with the mass and the emitted spectrum contains a larger number of $e^+e^-$. As the PBH mass increases, the Hawking spectrum becomes colder and the number of ionizing particles decreases, giving weaker bounds. For $M_{\rm PBH}\gtrsim 10^{17}\,{\rm g}$, the Hawking temperature falls below the electron mass. The emission of electrons and positrons is then exponentially suppressed and only arises from the high-energy tail of the thermal spectrum, making the ionization signal rapidly suppressed at larger masses.

The impact of diffusion is mostly visible at low PBH masses. In this region, the Hawking spectrum still contains a significant fraction of energetic particles, which can easily escape from the cloud before losing all their energy. Suppressing the diffusion coefficient confines these particles more efficiently increasing the ionization, thus the constraints. At larger PBH masses, the spectrum shifts to lower energies (See Appendix~\ref{sec:Setups}), remaining in the loss-dominated regime even for the largest value of $D_0$ tested here. As a result, the dependence on this value becomes very mild or negligible.

\begin{figure*}[t]
    \centering
     \includegraphics[width=0.495\linewidth]{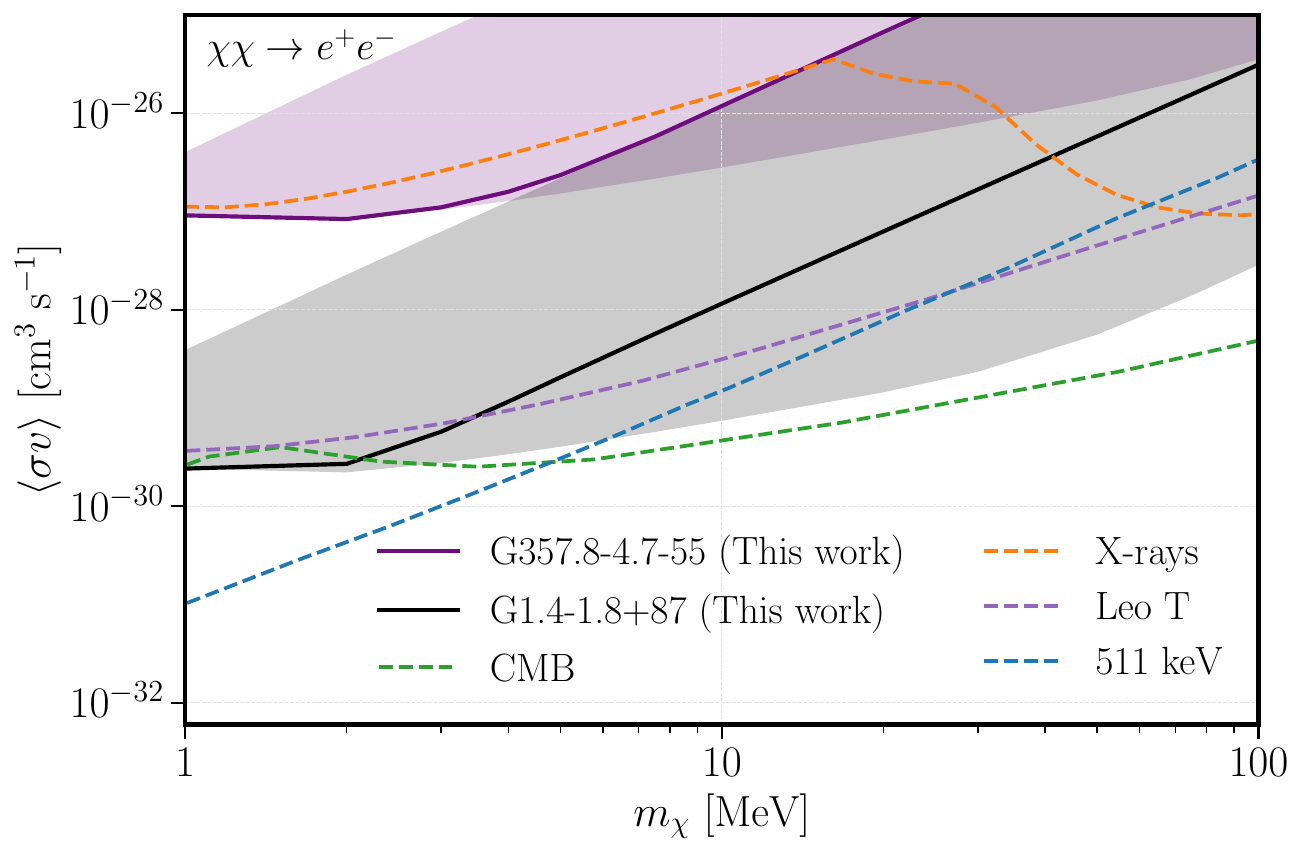}
    \includegraphics[width=0.495\linewidth]{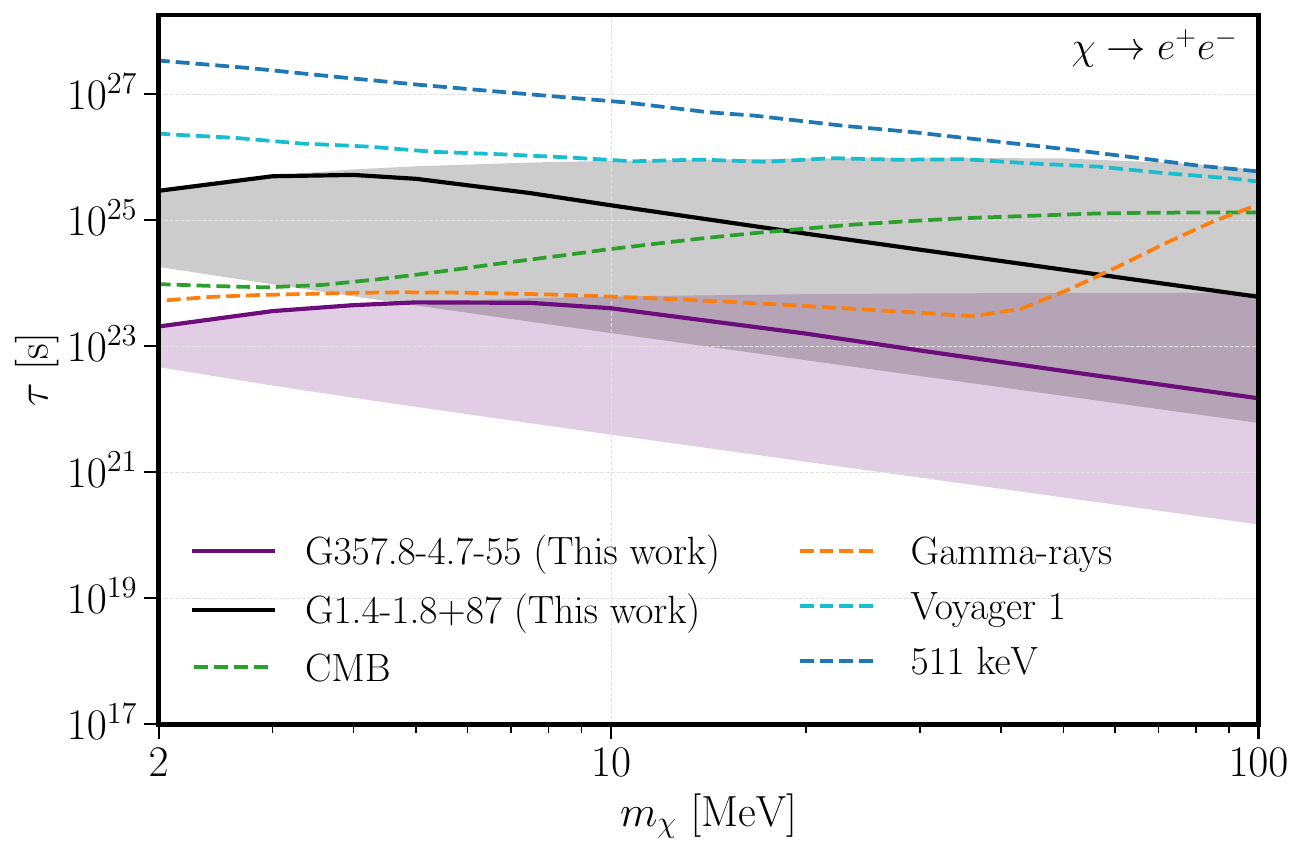}
    \includegraphics[width=0.53\linewidth]{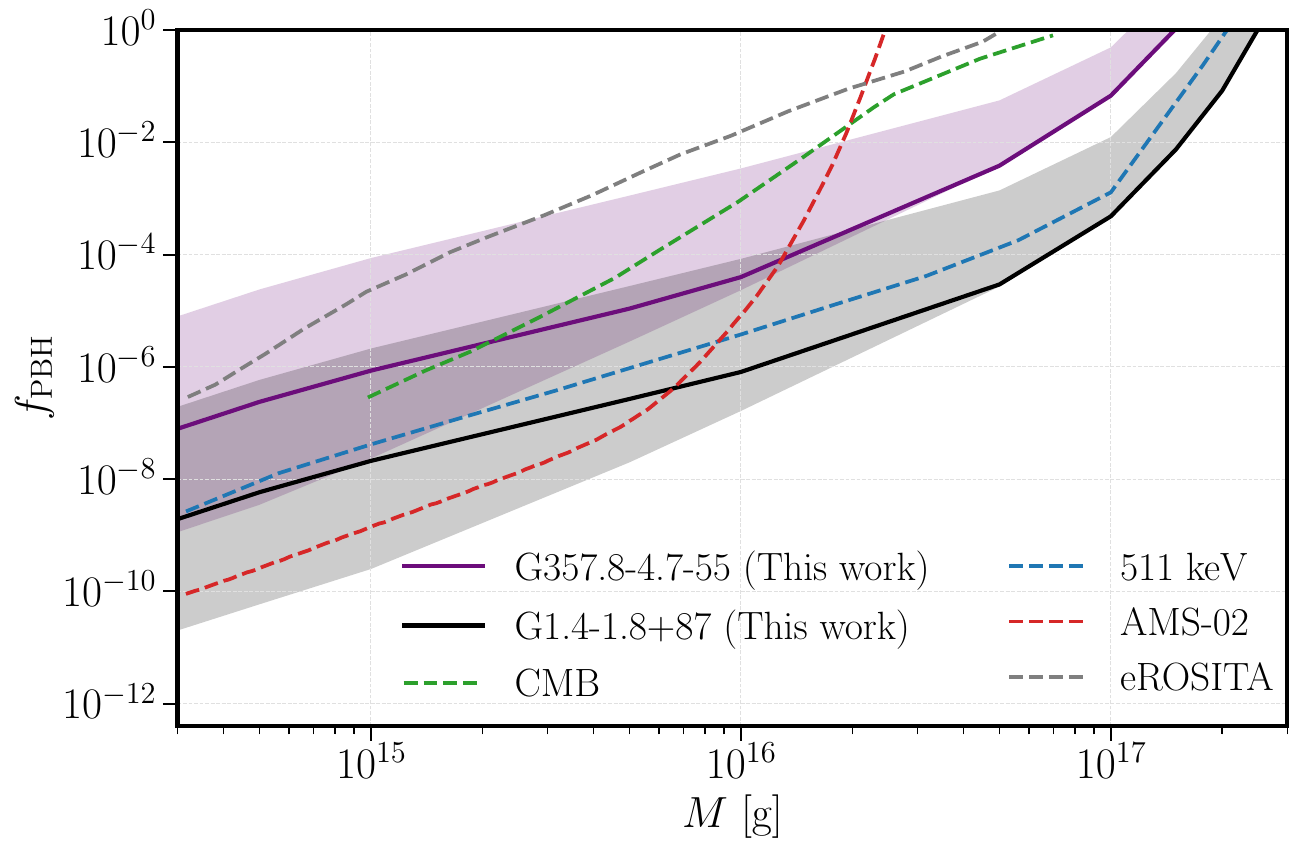}
    \caption{Comparison of our strongest clouds limits with existing constraints. \textbf{Top left:} annihilating DM constraints, compared with limits from the CMB (green)~\cite{Slatyer2016CMB}, Leo~T (purple)~\cite{Wadekar_2022}, X-rays (yellow)~\cite{Balaji:2025afr}, and the Galactic 511 keV line morphology (blue)~\cite{DelaTorreLuque2024}
    \textbf{Top right:} decaying DM constraints, compared with bounds from the CMB (green) and gamma rays (yellow) ~\cite{Essig:2013goa}, Voyager~1 (cyan) and the Galactic 511 keV line (blue) ~\cite{DelaTorreLuque2024}. \textbf{Bottom:} PBH constraints on the DM fraction, compared with existing limits from X-rays with eROSITA (grey)~\cite{Balaji:2025afr}, the Galactic 511 keV line (blue)~\cite{DelaTorreLuque2024}, AMS-02 (red)~\cite{Huang:2025AMS}, and CMB energy injection (green)~\cite{Liu2016CMB}.
}
    \label{fig:constcompare}
    \end{figure*}

We compare our strongest PBH constraints with existing limits in Fig.~\ref{fig:constcompare} (bottom panel), where we compare with limits derived from X-rays with eROSITA~\cite{Balaji:2025afr}, the 511 keV line~\cite{DelaTorreLuque2024}, AMS-02~\cite{Huang:2025AMS} and the robust CMB constraints~\cite{Liu2016CMB} (see also Ref.~\cite{CMB_Visinelli}). At lower PBH masses, our bounds remain weaker than AMS-02 constraints~\cite{Huang:2025AMS} and comparable to those coming from the 511 keV line~\cite{DelaTorreLuque2024}. However, above $\sim10^{16}\,\mathrm{g}$, the G1.4$-$1.8+87 cloud provides stronger limits than existing bounds, making it a particularly promising target for probing the PBH DM fraction in the upper end of this asteroid-mass window.

\section{Discussion and conclusions}
\label{sec:conclusions}

In this work, we have explored the ionization of cold gas clouds as a new probe of different DM scenarios producing low-energy electrons and positrons. Unlike conventional indirect searches that rely on the detection of particles escaping the production region, these objects provide an environment where sub-GeV particles can efficiently deposit their energy locally through ionization. This makes them particularly sensitive to light particle DM models in the MeV range and to asteroid-mass PBHs emitting low-energy electrons and positrons through Hawking radiation.

A central result of this work is that the sensitivity of the clouds to exotic ionization sources is governed by the interplay between the local DM density, the physical properties of the cloud, and particle propagation. We identify the most promising targets as large cold clouds with densities of a few tens of $\mathrm{cm}^{-3}$ to $\mathcal{O}(10^3)~\mathrm{cm}^{-3}$, which are able to retain a larger fraction of the energy injected by DM while being more effectively shielded from external ionization sources (CRs). Clouds located above the Galactic plane are expected to be even more favorable, as they are less exposed to CR-induced ionization and are therefore more sensitive to exotic ionization sources.
Our results also show that a higher DM density alone does not necessarily imply a better target. Although clouds closer to the Galactic Center experience an enhanced DM density, they are also subject to larger astrophysical ionization backgrounds, reducing the room for an additional exotic contribution. 

As we have shown, the propagation of charged particles inside cold clouds represents the main astrophysical uncertainty in this approach, at least for the targets studied in this work. The balance between energy losses and diffusion determines whether a cloud behaves approximately as a calorimetric target or whether a significant fraction of the injected particles escapes before depositing their energy. In the loss-dominated regime, the predicted ionization rate is relatively insensitive to the details of transport, while in the escape-dominated regime the constraints become strongly dependent on the assumed diffusion coefficient. This dependence highlights the importance of improving our understanding of charged-particle transport in these environments. This also implies that larger and denser clouds are less affected by uncertainties in the transport of charged particles. However, very dense clouds, with densities above $10^{4} \mathrm{cm}^{-3}$, significantly suppress the ionization rate expected from DM, leading to an optimal range of cloud densities around $\mathcal{O}(100),\mathrm{cm}^{-3}$.


An interesting point to note is that DM ionization produces a spatial ionization profile that is clearly distinguishable from that generated by standard CRs, providing a distinctive signature. CRs entering the cloud from the outside naturally produce ionization profiles enhanced near the cloud surface, as low-energy particles are progressively attenuated while penetrating into denser regions. In contrast, the DM contribution is generated throughout the volume and therefore leads to a smoother ionization profile, with a relative enhancement toward the cloud center. Consequently, a DM contribution would not simply modify the normalization of the standard ionization rate, but could also alter its spatial dependence. Spatially resolved measurements of molecular and atomic tracers could therefore provide an interesting way to distinguish exotic ionization sources from standard astrophysical backgrounds. We also note that it is possible a regime associated with very dense cores where propagation becomes ballistic. This, however, might affect dense cores, but not diffuse, extended clouds.

The homogeneous cloud approximation adopted in this work allows us to isolate the main physical effects and quantify the potential of MCs as DM targets. A more realistic description including density gradients, magnetic field configurations, and spatially varying diffusion coefficients will be required for a complete interpretation of individual clouds. 

Overall, we have shown that cloud ionization provides a complementary avenue to probe DM models that inject low-energy charged particles. This method is particularly promising for sub-GeV DM scenarios, where conventional searches can be limited by the challenges associated with detecting low-energy products. For MeV-scale annihilating DM, we find that the constraints obtained from MC ionization could even surpass those derived from the CMB anisotropy spectrum at low masses. Similarly, for PBH masses above $10^{16}\mathrm{g}$, the limits obtained from the cloud G1.4-1.8+87 surpass previous constraints even in conservative propagation scenarios, demonstrating the capability of these targets for probing the fraction of DM composed of PBHs below $10^{18}\mathrm{g}$.
A larger sample of cold clouds, particularly large near the Galactic Center and located above the Galactic plane with reduced astrophysical backgrounds, together with improved modeling of CR transport, could substantially enhance the sensitivity of this approach.

\section*{Acknowledgments}
We thank Zhaodong Shi for sharing his code with us, which provided a very useful starting point for this analysis. We also thank Rebecca Leane for her helpful comments and suggestions.
ASP acknowledges support from a European traineeship programme after master studies funded through Erasmus+, which enabled a research stay in Stockholm, and thanks Tim Linden and his group for their hospitality during this visit.
PDL has been supported by the Juan de la Cierva JDC2022-048916-I grant, funded by MCIU/AEI/10.13039/501100011033 European Union "NextGenerationEU"/PRTR, and is currently supported by Ramón y Cajal RYC2024-048445-I grant, which is funded by MCIU/AEI/10.13039/501100011033 and FSE+. The work of PDL is also supported by the grants PID2021-125331NB-I00 and CEX2020-001007-S, both funded by MCIN/AEI/10.13039/501100011033 and by ``ERDF A way of making Europe''. PDL also acknowledges the MultiDark Network, ref. RED2022-134411-T. This project used computing resources from the National Academic Infrastructure for Supercomputing in Sweden (NAISS) under projects NAISS 2025-5-729 and NAISS 2024/5-666.

\appendix
\onecolumngrid
\vspace{2cm}

\section{Additional plots}
\label{sec:Setups}

In this appendix we show additional plots that complement the discussion in the main text. These figures show the propagated electron and positron spectra and the corresponding ionization profiles for different source models in the same targets. 

\begin{figure*}[t]
    \centering

    \subfloat[Annihilating DM]{%
    \includegraphics[width=0.49\textwidth]{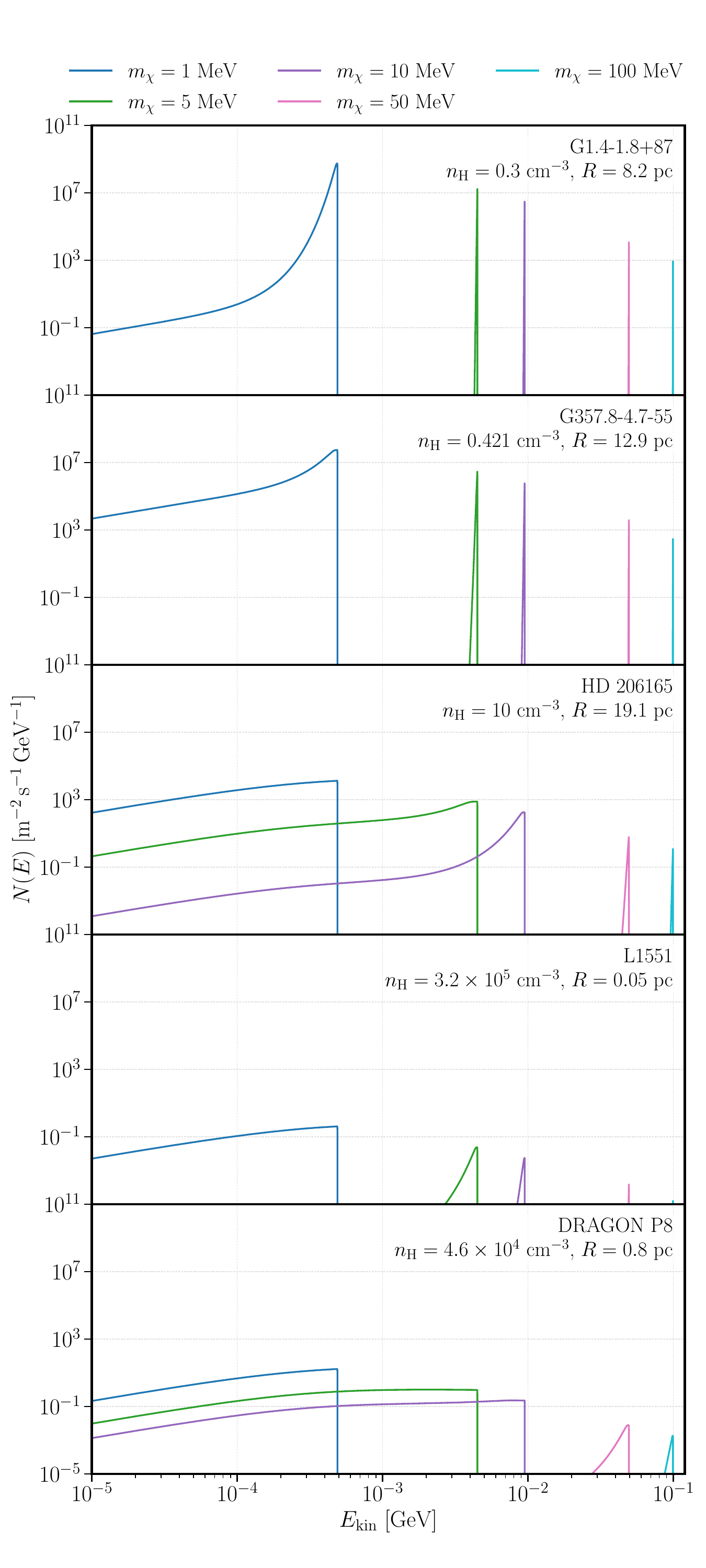}
    }
    \hfill
    \subfloat[PBHs]{%
    \includegraphics[width=0.49\textwidth]{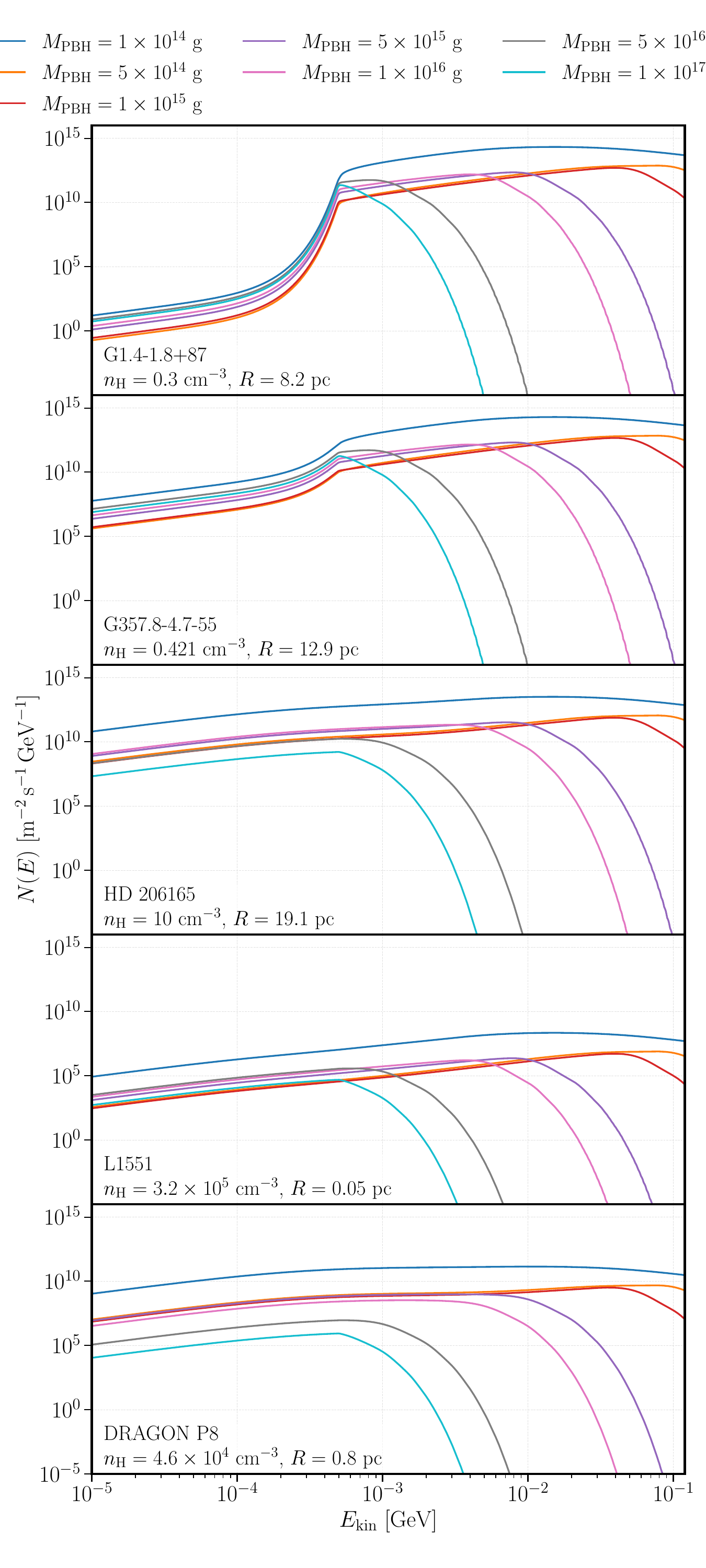}
    }

    \caption{
    Propagated electron and positron spectra inside the target clouds for: \textbf{(a)} annihilating DM with $\langle \sigma v \rangle = 1\times10^{-30} {\rm cm^{3}s^{-1}}$ and \textbf{(b)} PBHs with $f=1$. Each panel corresponds to a different cloud, with the adopted gas density $n_{\rm H}$ and radius $R$ indicated in the plot. The different colors show the stationary spectra obtained for various DM or PBH masses for a fixed diffusion setup with $D_0=3\times10^{28}\,{\rm cm^2\,s^{-1}}$. The peaks in the annihilating-DM case correspond to the monoenergetic injection energy, while the low-energy tail is produced by energy losses during propagation. In the PBH case, the injected spectrum is continuous because it originates from Hawking evaporation.
    }
    \label{fig:multicloud_spectra}
\end{figure*}

Figure~\ref{fig:multicloud_spectra} shows the steady-state spectra obtained inside L1551-IR core and G1.4-1.8+87 for annihilating DM and evaporating PBHs. An interesting feature of the PBH case is the shape of the propagated spectrum. When energy losses dominate, the low-energy part of the spectrum remains relatively flat, since particles lose most of their energy inside the cloud. In contrast, sharp drops at higher energies indicate that these particles escape efficiently before losing a significant fraction of their energy.

\begin{figure*}[t]
    \centering

    \subfloat[L1551-IR core, decaying DM]{%
        \includegraphics[width=0.49\textwidth]{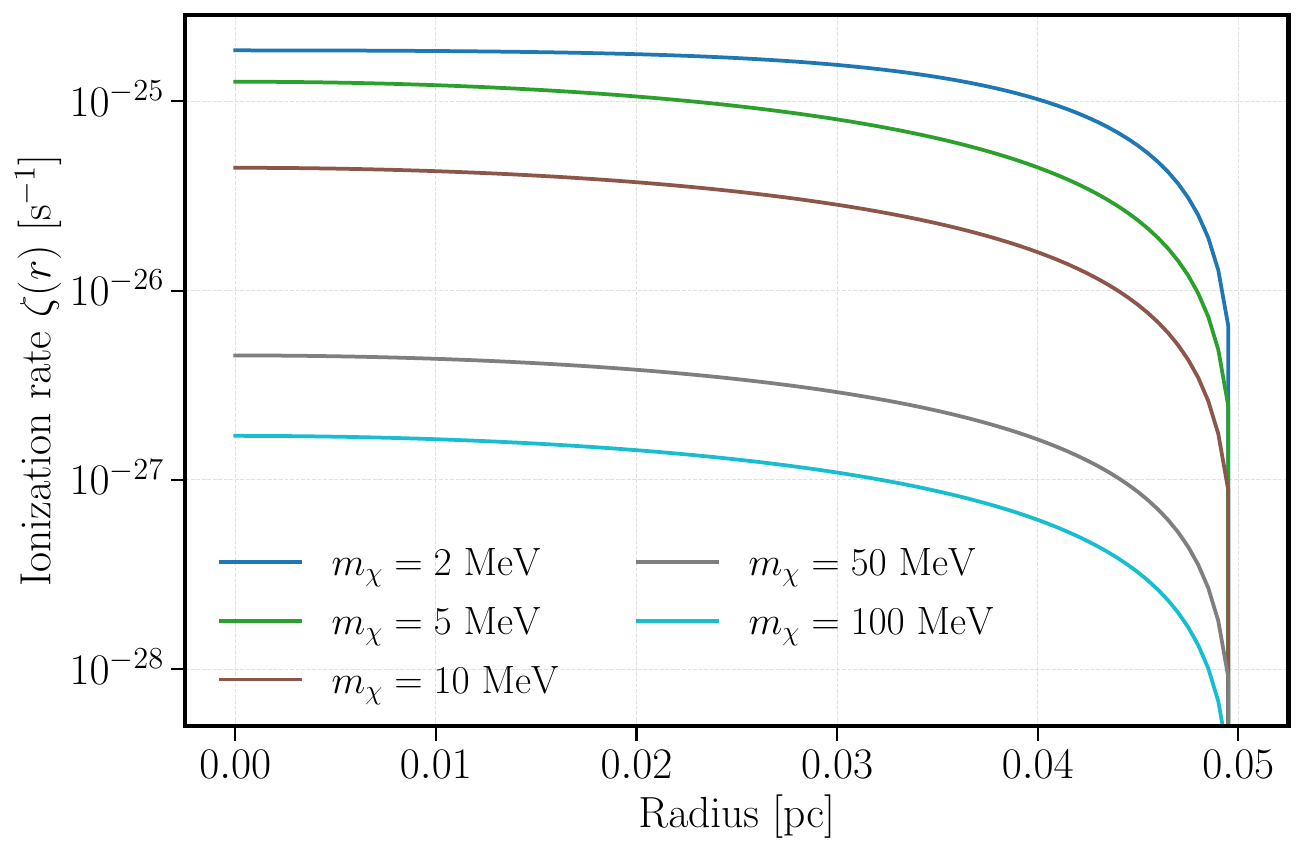}
    }
    \hfill
    \subfloat[L1551-IR core, PBHs]{%
        \includegraphics[width=0.49\textwidth]{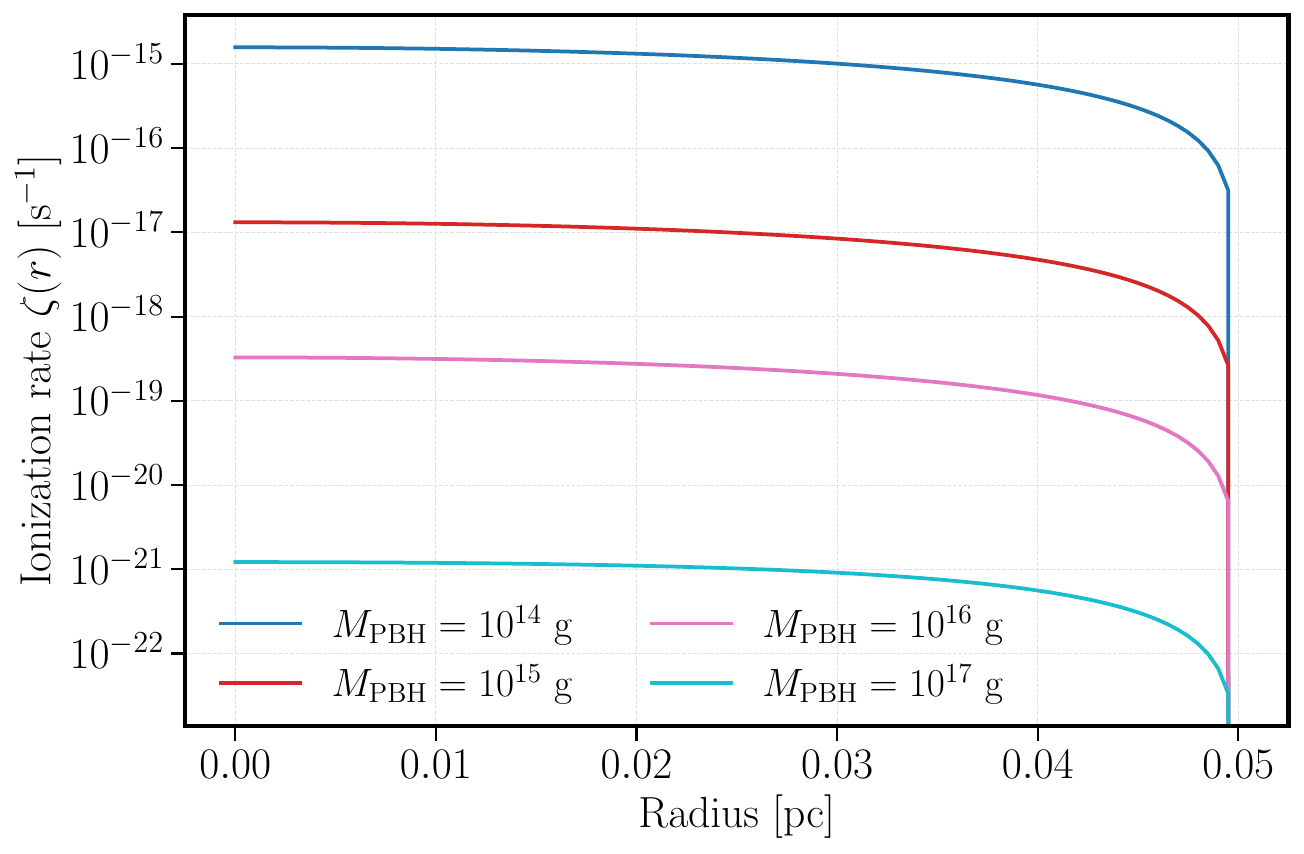}
    }

    \vspace{0.3cm}

    \subfloat[G1.4-1.8+87, decaying DM]{%
        \includegraphics[width=0.49\textwidth]{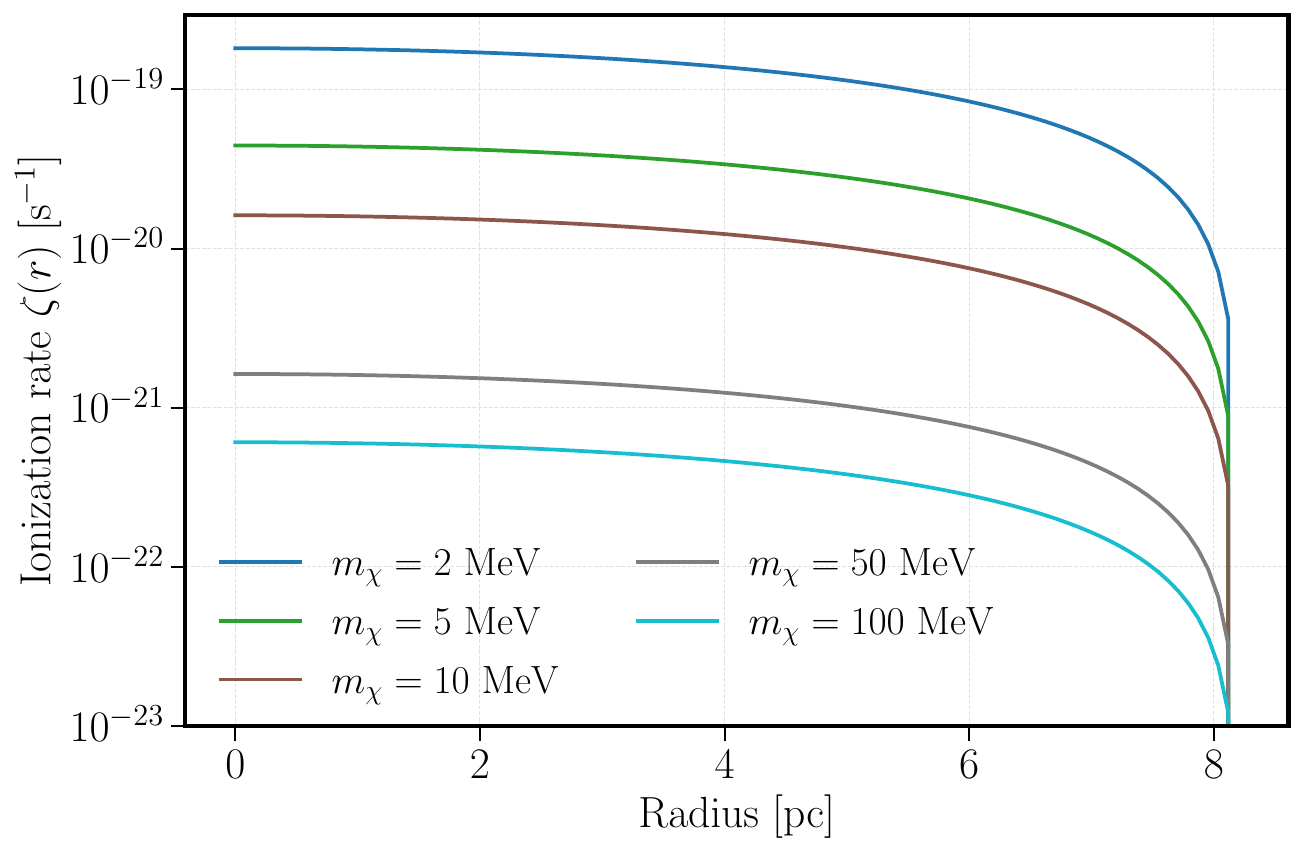}
    }
    \hfill
    \subfloat[G1.4-1.8+87, PBHs]{%
        \includegraphics[width=0.49\textwidth]{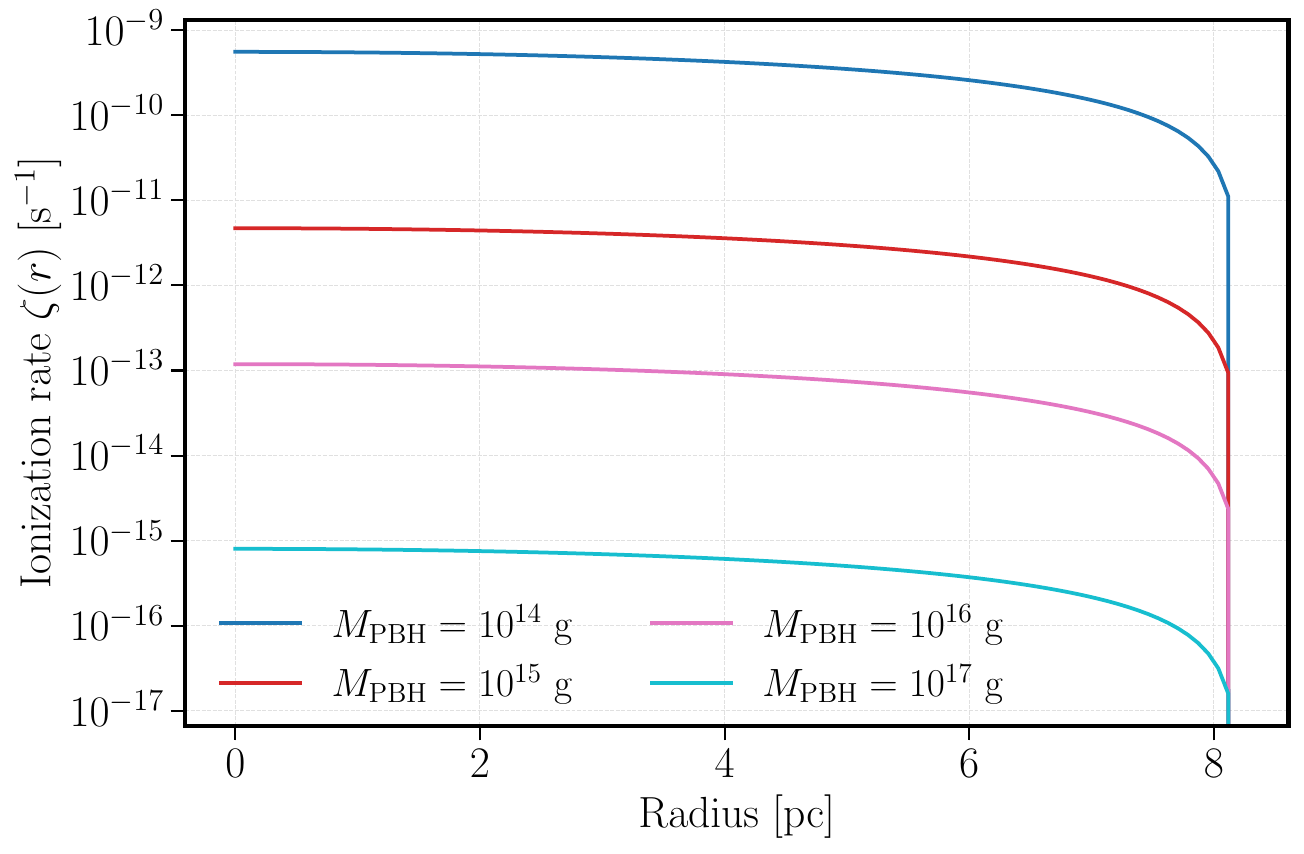}
    }

    \caption{
    Ionization profiles induced by decaying DM (left panels) and evaporating PBHs (right panels) in the L1551-IR core and in G1.4-1.8+87. The top panels correspond to the     L1551-IR core, while the bottom panels correspond to G1.4-1.8+87. The left  panels show decaying DM with $\tau=1\times10^{26}\,{\rm s}$ for different DM masses, and the right panels show the PBH contribution for different PBH masses with $f=1$. In all cases, the diffusion normalization is fixed to $D_0=3\times10^{28}\,{\rm cm^2\,s^{-1}}$.
    }
    \label{fig:ionizprofiles_decay_pbh}
\end{figure*}

Figure~\ref{fig:ionizprofiles_decay_pbh} shows the corresponding radial ionization profiles for decaying DM and PBHs in L1551-IR and G1.4-1.8+87. These examples illustrate how the same injected spectrum can lead to different profiles and total ionization rates depending on the cloud size and gas density.

\section{Effect of the surrounding gas in L1551-IR core}
\label{App:L1151}

In the Sec.~\ref{sec:LISIon} we studied the LIS-induced ionization in L1551-IR, and modelled the target as an isolated dense core. However, the core is embedded in the larger L1551 MC, and the surrounding gas may affect the penetration of low-energy CRs. To estimate the extent of this effect, we compare the naked-core setup with a core + cloud model, in which the dense L1551-IR is within a diffuse envelope with $n_{\rm H}=200\,{\rm cm^{-3}}$, connected to the core through a smooth transition.

\begin{figure}[h]
    \centering
    \includegraphics[width=0.6\linewidth]{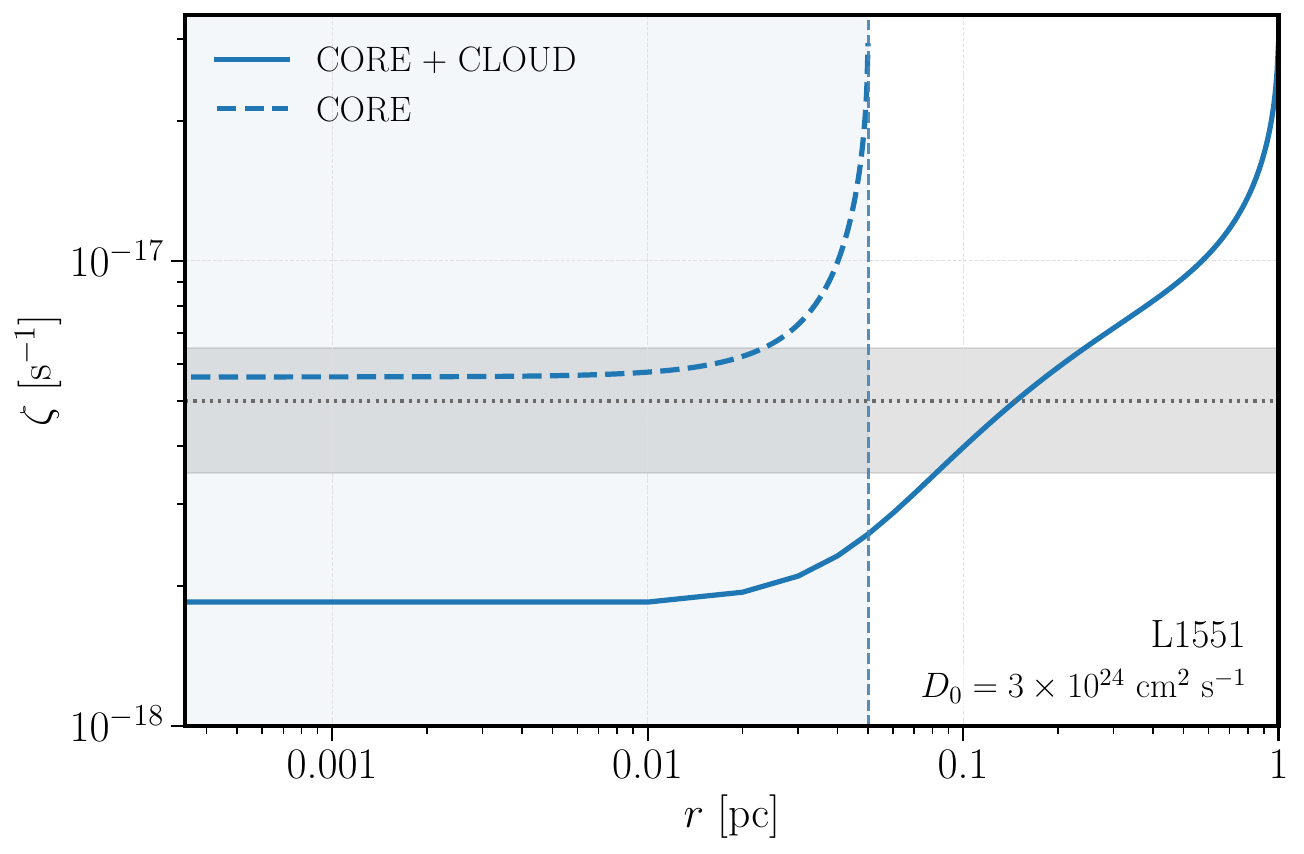}
    \caption{Comparison of the expected ionization rate when including a diffuse gas around the L1551 core and when using the naked core. The blue dashed vertical line shows the radius of the core that is connected to the diffuse environment with $n_H = 200 {\rm cm^{-3}}$ through a smooth transition. The horizontal dotted line represents the measured ionization rate with the grey band representing its uncertainty.}
    \label{fig:corepluscloud}
\end{figure}

The comparison is shown in Fig.~\ref{fig:corepluscloud}. We use the suppressed-diffusion benchmark, $D_0=3\times10^{24},{\rm cm^2,s^{-1}}$, since this is the regime required to reproduce the low ionization rate inferred for the target. Including the surrounding diffuse gas reduces the predicted ionization rate in the core by a factor of $\sim3$ compared to the naked-core calculation. This difference can be largely compensated by increasing the diffusion coefficient to $\sim4\times10^{24},{\rm cm^2,s^{-1}}$. However, this treatment depends on how the transition between the dense core and the diffuse cloud is modeled, as well as on the assumed properties of the extended diffuse cloud, both of which remain poorly constrained observationally and therefore introduce additional uncertainties.

For this reason, we adopt the simpler naked-core setup in the main analysis. Using the full cloud model instead would not significantly modify the reduced-diffusion bands in our constraints. For such strongly suppressed diffusion, energy losses already dominate over particle escape for most of the relevant clouds and particle energies. Consequently, the predicted ionization rates remain very similar, and our main conclusions are unchanged.

\section{DM profile uncertainties}

\begin{figure}[H]
    \centering
    \includegraphics[width=0.495\linewidth]{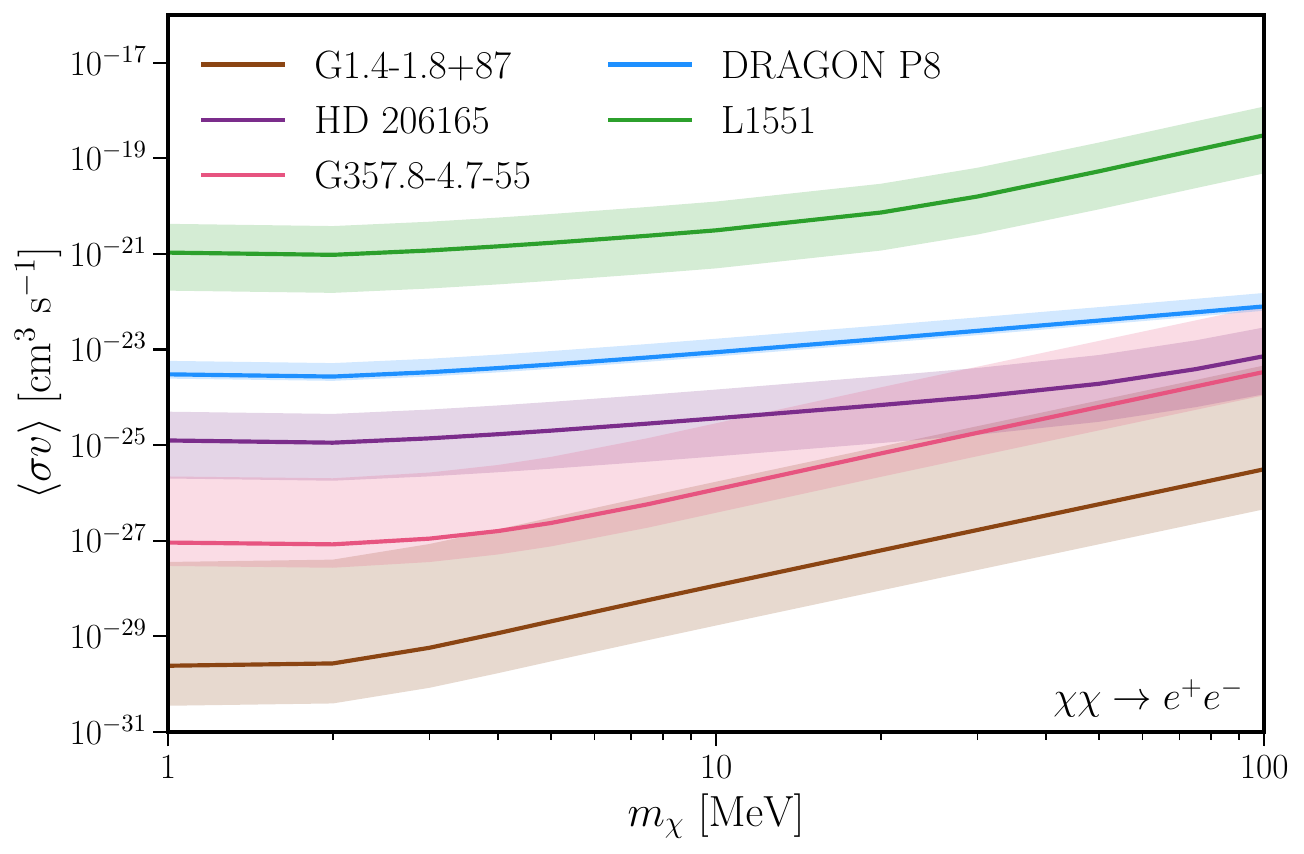}
    \includegraphics[width=0.495\linewidth]{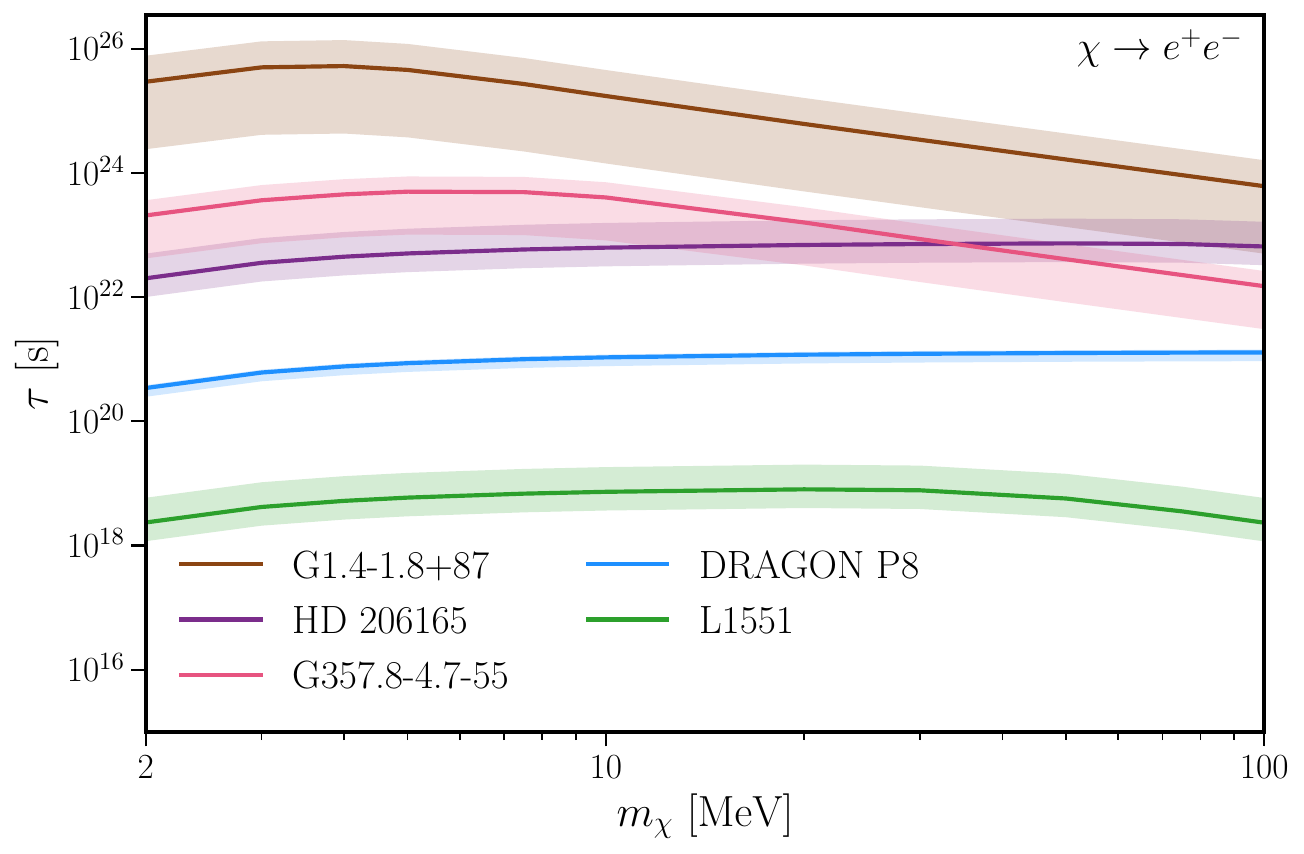}
    \includegraphics[width=0.55\linewidth]{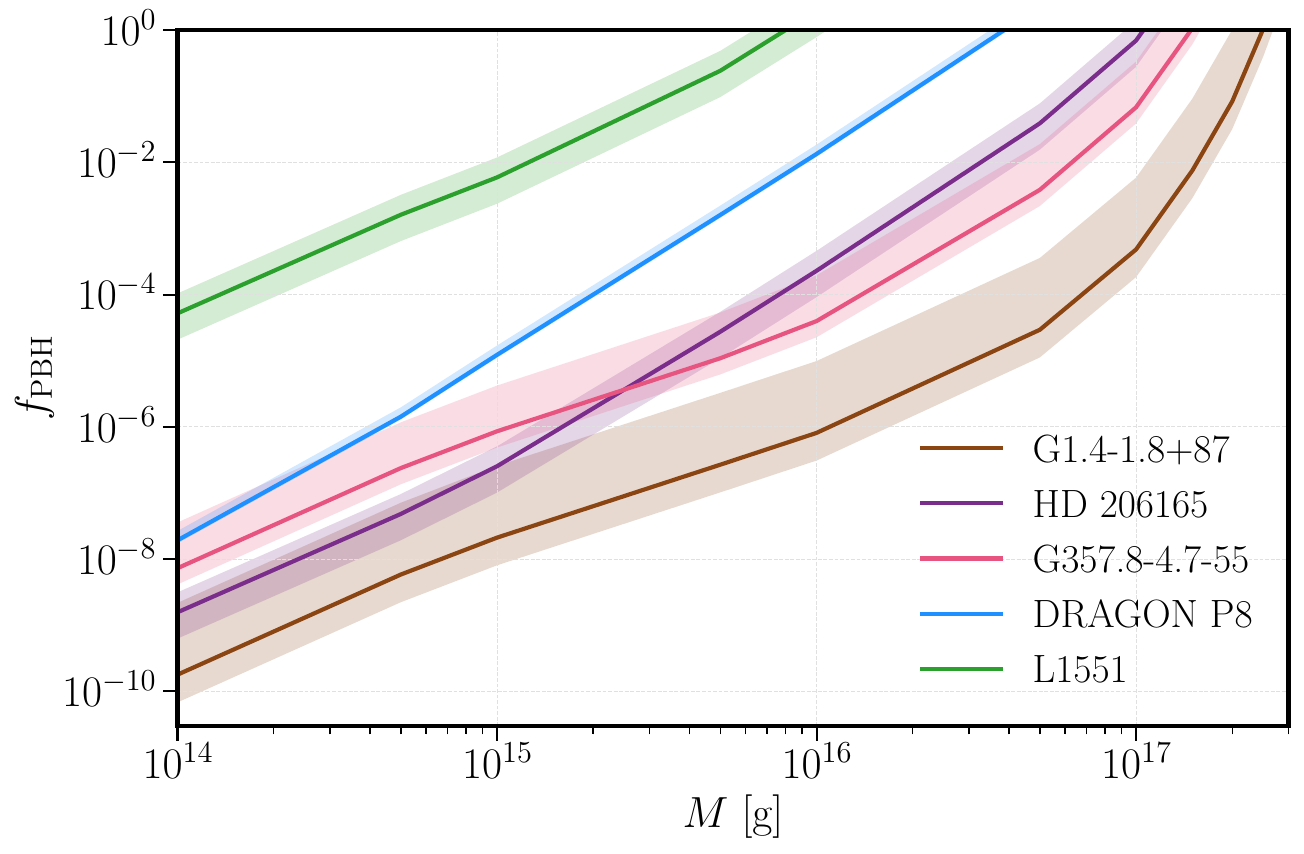}
    \caption{DM constraints for different assumptions of the Galactic DM density profile. The variation between the curves illustrates the impact of uncertainties in the DM distribution on the derived limits. The case of annihilating DM is represented in the top-left plot, decaying DM in the top-right plot and PBHs in the bottom panel.}
    \label{fig:constprofiles}
    \end{figure}

The constraints are also affected by the uncertainty in the Galactic DM density profile. This can be seen in Fig.~\ref{fig:constprofiles}. The effect is more relevant for clouds close to the Galactic Center, since they are more sensitive to the DM distribution. The uncertainty is larger in the case of annihilating DM because the signal scales as $\rho_\chi^2$ while for decaying DM and PBHs only as $\rho_\chi$. We quantify this by comparing the benchmark NFW distribution with Moore and Burkert profiles as optimistic and pessimistic cases. For the local clouds, we adopt a reference local DM density of $\rho_\odot = 0.4\,{\rm GeV\,cm^{-3}}$~\cite{Benito_2019} and vary it within the range $\rho_\odot=0.2$--$1\,{\rm GeV\,cm^{-3}}$ to assess the impact of this uncertainty.

\label{sec:profileunc}

\section{Comparison of e-H ionization cross sections}
\label{sec:crosssectiondif}
\begin{figure}[tb]
    \centering
    \includegraphics[width=0.5\linewidth]{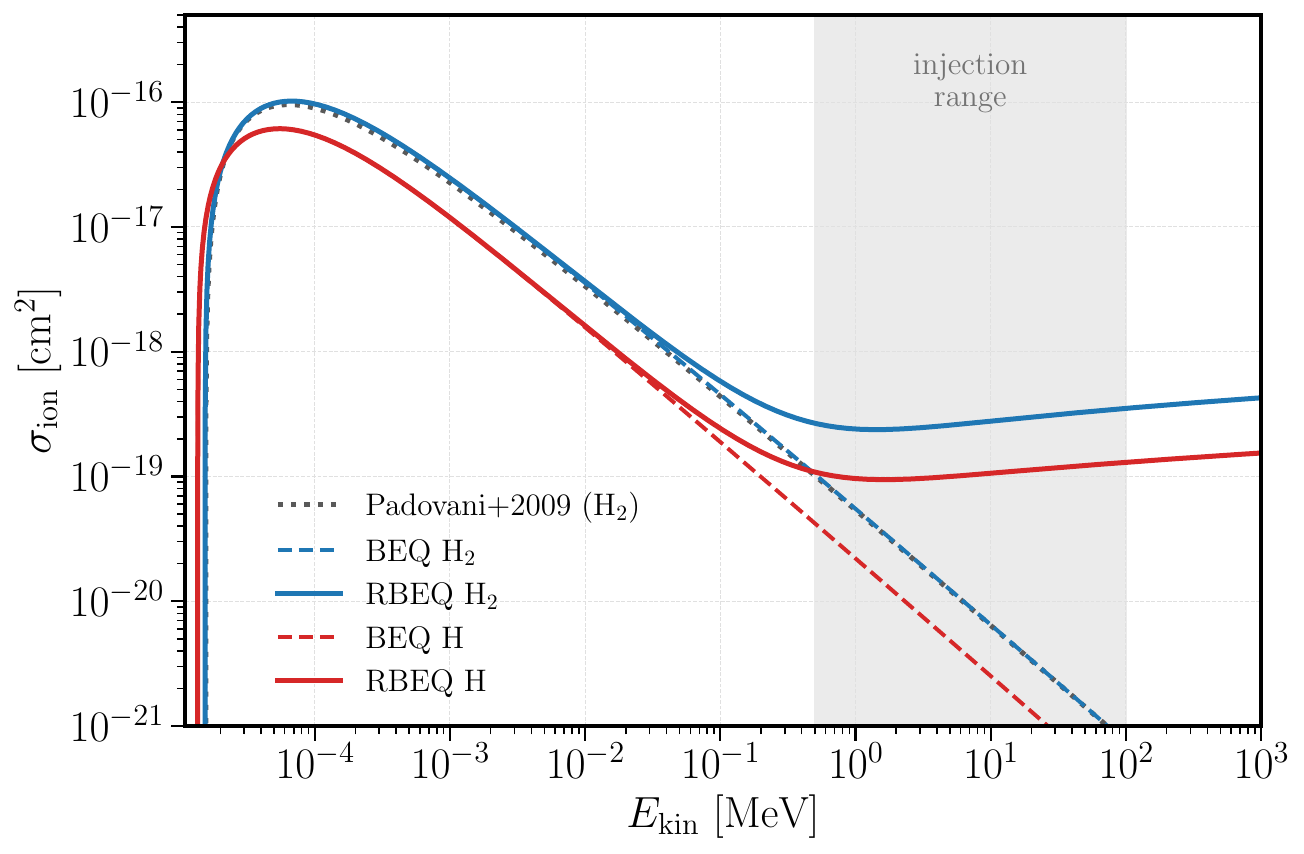}
    \caption{Electron-impact ionization cross sections for molecular and atomic hydrogen. The dotted grey curve shows the parametrization adopted in Ref.~\cite{Padovani:2009bj}, while the dashed and solid curves show the non-relativistic~\cite{PhysRevA.50.3954} and relativistic~\cite{PhysRevA.62.052710} binary-encounter prescriptions, respectively, for $\mathrm{H}_2$ and $\mathrm{H}$. The shaded region indicates the range of injection energies relevant for the DM and PBH scenarios considered in this work.}
    \label{fig:xsec}
\end{figure}

The ionization rates derived in this work depend directly on the electron ionization cross section. In our case, the electron energies relevant for our analysis extend up to $\sim100$~MeV, since we are probing electrons injected with sub-GeV energies. At these energies, relativistic corrections become very important. We therefore compare in Fig.~\ref{fig:xsec} the semi-empirical parametrization adopted in Ref.~\cite{Padovani:2009bj} (Padovani+ 2009) with the binary-encounter models of Kim and Rudd~\cite{PhysRevA.50.3954} (BEQ) and their relativistic extension \cite{PhysRevA.62.052710} (RBEQ). These are shown for both, interaction of electrons with atomic and molecular hydrogen, which leads to an ionization difference by a factor slightly lower than 2~\cite{CloudyHazyC23,Liszt:2002xa, Indriolo:2007nv}. The different prescriptions agree well at low energies but begin to differ significantly in the MeV regime, where relativistic effects become important.

As discussed in Ref.~\cite{PhysRevA.62.052710}, relativistic effects approximately double the ionization cross section of atomic hydrogen already at $T\simeq300$~keV and become increasingly important at higher energies. We recover the characteristic flattening and subsequent rise of the relativistic cross section, with a minimum around the MeV scale. This behavior is qualitatively analogous to the relativistic rise familiar from Bethe-type stopping powers: as $\beta\rightarrow1$, the velocity-dependent suppression saturates while the logarithmic relativistic contribution continues to increase.

\begin{figure}[H]
    \centering
    \includegraphics[width=0.495\linewidth]{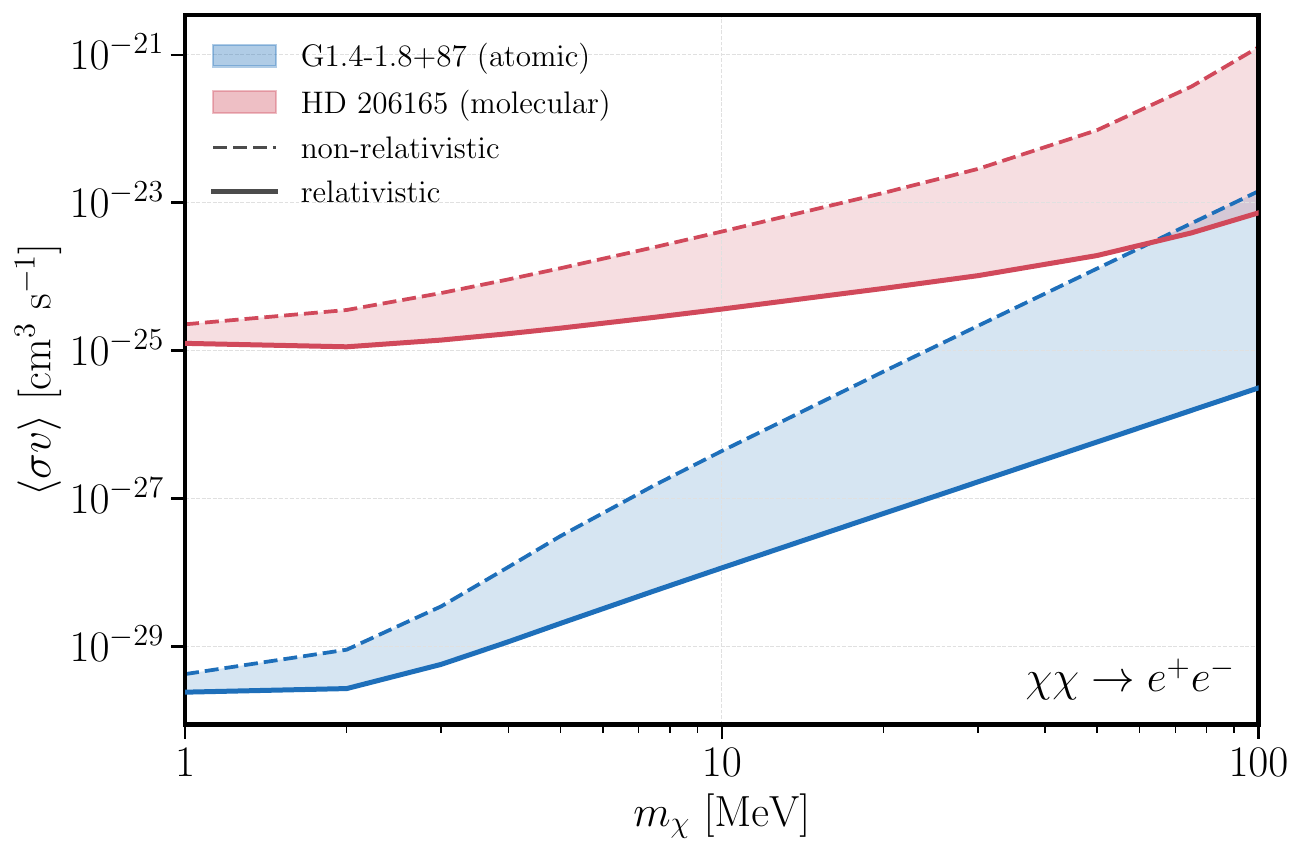}
    \includegraphics[width=0.495\linewidth]{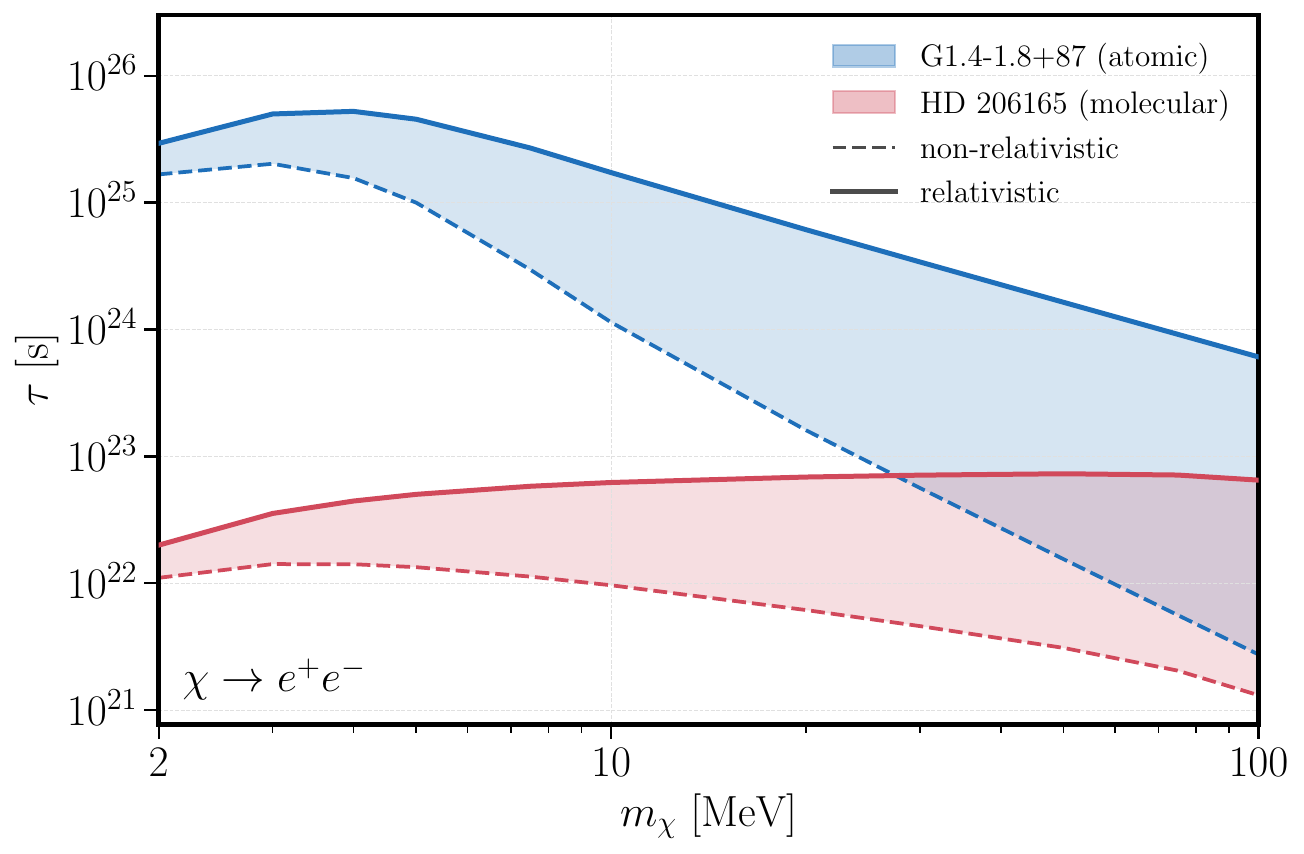}
    \includegraphics[width=0.55\linewidth]{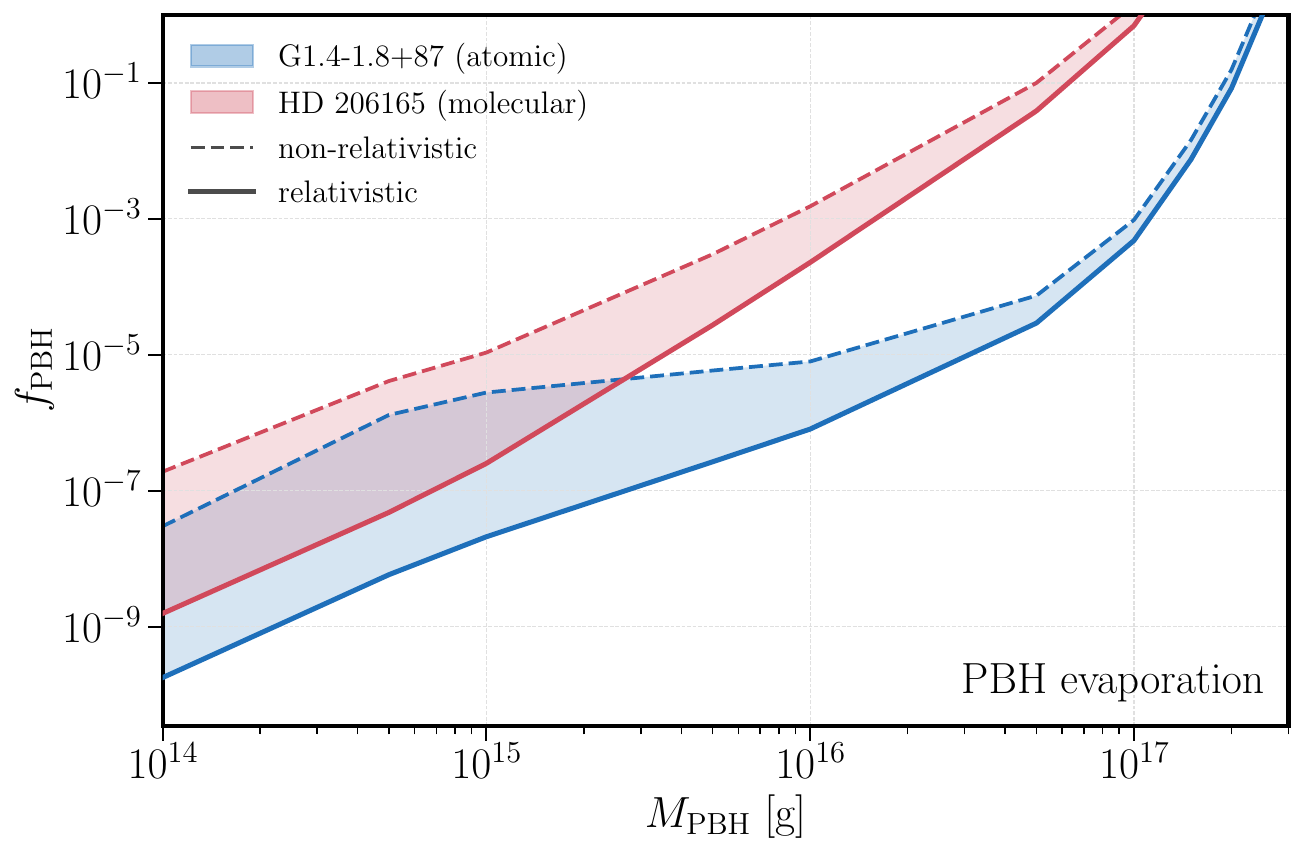}
    \caption{Impact of the treatment of the ionization cross section on the derived constraints, for an atomic cloud (G1.4-1.8+87, blue) and a molecular one (HD~206165, red). Dashed lines use the non-relativistic cross sections (BEQ for atomic hydrogen~\cite{PhysRevA.50.3954} and the Padovani et al. parametrization for H$_2$~\cite{Padovani:2009bj}), solid lines the relativistic RBEQ treatment of Kim et al.~\cite{PhysRevA.62.052710}, and the shaded band is the parameter space gained. Annihilating DM is shown in the top-left plot, decaying DM in the top-right plot and PBHs in the bottom panel.}
    \label{fig:tripanelsigma}
\end{figure}

Adopting the relativistic parametrization appreciably changes the total expected ionization, as shown in Fig.\ref{fig:tripanelsigma} for two of our clouds. 
As one can see, the predicted ionization rates increase for the more energetic injection spectra, leading to stronger constraints. The effect becomes particularly important at large DM masses for both annihilation and decay, and at low PBH masses, for which the Hawking spectra extend to higher electron and positron energies.

\newpage

\bibliographystyle{apsrev4-1}
\bibliography{references.bib}

\end{document}